\documentclass[aps,amsmath,amssymb,floatfix,superscriptaddress,nofootinbib,a4paper]{revtex4}

\usepackage{graphicx}
\usepackage{hyperref}

\allowdisplaybreaks[4]

\renewcommand{\Re}{\operatorname{Re}}
\renewcommand{\Im}{\operatorname{Im}}

\begin{document}
\title{%
On the non-hermitian Kitaev chain
}
\author{Eddy Ardonne}
\email[Corresponding author: ]{ardonne@fysik.su.se}
\affiliation{Department of Physics, Stockholm University, AlbaNova University Center, 106 91 Stockholm, Sweden}
\author{Viktor Kurasov}
\affiliation{Department of Physics, Stockholm University, AlbaNova University Center, 106 91 Stockholm, Sweden}
\begin{abstract}
We study the non-hermitian Kitaev chain model, for arbitrary complex parameters. In particular, we give a concise
characterisation of the curves of eigenvalues in the complex plane in the infinite size limit, using a novel method
which can be applied to other non-hermitian systems. 
Using this solution, we characterise under which conditions
the skin effect is absent, and for which eigenstates this is the case. We also fully determine the region in parameter
space for which the model has a zero mode.
\end{abstract}
\date{\today}
\maketitle

\section{Introduction}

The study of non-hermitian systems, in various contexts, has been extremely intense
during the last years. These studies cover the properties of non-hermitian systems in general
(focussing on the differences and similarities with hermitian systems), studies of particular
non-hermitian models, as well as utilising non-hermitian systems for actual applications, using
various different types of physical systems
\cite{
AgKoLaSe2024,
AgKoLaSe2023unpub,
BeBuKu2019,
BrLiBeRoAr2022unpub,
Br2014,
DeYoHa2021,
Ed2022,
EdKuBe2019,
Ez2019,
Ez2024,
GoAsKaTaHiUe2018,
KaBeSa2019,
KaShUeSa2019,
KoBu2020,
KuDw2019,
KuEdBuBe2018,
Ma2024a,
Ma2024b,
MaBe2021,
MaRo2023,
MaVaTo2018,
LaQv2023,
Le2016,
LeLiGo2019,
LeBlHuChNo2017,
LiWaDuCh2024unpub,
LiWeWuRuChLeNi2024unpub,
LiLi2024,
LiTaYaLiLe2023,
LiLiAn2023,
MaAuPePaAc2024,
MoArBe2023,
NeRo2024,
OkKaShSa2020,
OrHe2023,
RaSiRaJi2024unpub,
RiRo2022,
RoOrAgHe2024,
SaCiCABa2024,
ShZhFu2018,
StBe2021,
VyRo2021,
YaBe2024unpub,
YaMa2023,
YaMoBe2023,
YaWa2018,
YoPeKaHa2019,
XiDeWaZhWaYiXu2019,
Xi2018,
ZeBe2024,
ZhSoZhZhLiCaLi2024,
ZhLe2019}.
One aspect that is of particular relevance for the current paper is the presence of the skin-effect, which
is closely related to the breakdown of the famed bulk boundary correspondence of hermitian topological systems
\cite{
HeBaRe2019,
Lo2019,
LeTh2019,
LiZhAiGoKaUeNo2019,
FlZoVaBeBaTi2022,
ZiReRo2021,
BoKrSl2020,
YoMu2019,
YaMoBe2021,
EdKuYoBe2020,
Sc2020,
BrHy2019,
EdAr2022}.

In this paper, we focus solely on Kitaev chain \cite{Ki2001} famous for its Majorana zero modes in its topological phase
(see \cite{Al2012} for a review).
We consider the non-hermitian version of this model, which has, in various incarnations, been studied before
\cite{
Yu2017,
KlCaDaMaWu2017,
LiZhZhSo2018,
LiJiSo2020,
KaAsKaUe2018,
Li2019,
YaSo2020,
ZhGuKoZhLi2021,
SaNiTa2022,
SaLa2023,
ShSo2023,
LiWaSoWa2024,
RaRoKuKaSa2023,
Zh2024,
MoNa2024,
FuKaMo2024,
PiMeAg2024unpub,
CaAg2024unpub,
LiLiZhTiArLi2024}.
A main result is that we concisely characterise the location of the eigenvalues in the complex plane in
the limit for an open chain in the infinite system size. Using our novel method, we
can explain why the curves the eigenvalues lie on exhibit branch points. Using our solution, we can address several
other interesting questions.

It is clearly natural to ask for which general choice of parameters does the
non-hermitian Kitaev chain have a zero-mode. Interestingly, the full answer to this question is
arguably a bit more complicated than one might expect.

The eigenvalues of generic non-hermitian systems are known to be very sensitive to the boundary conditions
\cite{KuEdBuBe2018,KoBu2020}.
This is due to the skin effect. Eigenstates exhibiting the skin effect are localised at the boundary of the system,
with an amplitude that decays exponentially in the bulk of the system.
Because of this exponential localisation of the eigenstates near the boundary, it is clear that the system
will be sensitive to small changes in the boundary conditions. In addition, the usual algorithms to obtain the eigenvalues
numerically becomes unstable for large system sizes.
For one-dimensional systems, one can use knowledge of the periodic system in order to predict whether or not
the open system exhibit skin-effect or not. In particular, one considers the eigenvalues of the periodic system, and
determines if this winds around an arbitrary point \cite{GoAsKaTaHiUe2018,KaShUeSa2019,OkKaShSa2020}.
If such a winding exists, the system exhibits a skin effect. It is, however, not a priori clear for which
choices of parameters in the model this occurs, and if it occurs, to which eigenvalues it pertains. 
Because of this, it is interesting to find an exact solution of the non-hermitian model one studies.
By this we mean a concise characterisation of the eigenvalues, that can be easily solved numerically, without
stability issues.

In the current paper, we use our novel method to obtain the eigenvalues in the thermodynamic limit in order to
provide explicit answers to the questions we posed above for the non-hermitian Kitaev chain.
We introduce the non-hermitian Kitaev chain, in order to set the notation, and start with the general analysis to solve
the model in Sec.~\ref{sec:model}.
We continue in Sec.~\ref{sec:finite-size} by providing
the solution of the non-hermitian Kitaev chain, for {\em chains of finite length}, under the
restriction that the left and right hopping parameters $t_1$ and $t_2$ are equal.
In Sec.~\ref{sec:zero-modes} we give a complete characterisation of the parameters, for which the
(non-hermitian) model has a zero-mode.
Here, we directly work in the infinite system size limit.
In Sec.~\ref{sec:Linfinity}, we provide a concise characterisation of the eigenvalues of the model,
again for fully generic parameters, in the infinite system size limit.
Using this characterisation, one can easily
obtain the curves in the complex plain corresponding to the eigenvalues of the model.
In Sec.~\ref{sec:skin-effect}, we analyse the solution to determine under which conditions the model
exhibits skin effect, and if this is the case, to which eigenvalues this pertains.
In Sec.~\ref{sec:summary} we summarise our results by means of an example, followed by a short
discussion in Sec.~\ref{sec:discussion}.

\section{The non-hermitian Kitaev chain}
\label{sec:model}
In this section we state the model we are interested in, namely the Kitaev chain \cite{Ki2001} with
nearest neighbour hopping and pairing, for arbitrary, complex parameters and start with the
analysis to find the eigenvalues. 
We are mainly interested in open chains, with `free' boundary conditions. In this paper, we do not study
the transition from the open to the periodic system. In terms of fermion creation and annihilation operators,
$c_j^\dagger$, $c_j$, the model for $L$ sites reads 
\begin{equation}
\label{eq:nh-top-sc}
H =
\sum_{j=1}^{L} \frac{m}{2} \bigl( c_j^\dagger c_j - c_j c_j^\dagger \bigr)
+ \sum_{j=1}^{L-1} \Bigl( t_1 c_j^\dagger c_{j+1} + t_2 c_{j+1}^\dagger c_j
+ d_1 c_j^\dagger c_{j+1}^\dagger + d_2 c_{j+1} c_{j} \Bigr) \ .
\end{equation}
If the model were hermitian, $m$ would take the role of the chemical potential, $t_1$ and $t_2$ would be the hopping parameters, while
$d_1$ and $d_2$ would correspond to the superconducting order parameter. 

To analyse the model, we write the hamiltonian in Bogoliubov-de Gennes form, that is, we write
\begin{equation}
H = \frac{1}{2} \Psi^\dagger \cdot \mathcal{H}_{\rm BdG} \cdot \Psi \ ,
\end{equation}
where $\Psi^\dagger = (c_1^\dagger, c_1, c_2^\dagger , c_2, \ldots )$.
For the model Eq.~\eqref{eq:nh-top-sc}, we find that $\mathcal{H}_{\rm BdG}$ is the following
$2L \times 2L$ matrix
\begin{equation}
\label{eq:nh-top-sc-matrix}
\mathcal{H}_{\rm BdG} =
\begin{pmatrix}
m & 0 & t_1 & d_1 & 0 & 0 & \cdots & \cdots & 0 & 0\\
0 & -m & -d_2 & -t_2 & 0 & 0 & \cdots & \cdots & 0 & 0 \\
t_2 & -d_1 & m & 0 & t_1 & d_1 & & & \vdots & \vdots \\ 
d_2 & -t_1 & 0 & -m & -d_2 & -t_2 & & & \vdots & \vdots \\ 
0 & 0 & t_2 & -d_1 & m & 0 & \ddots & & 0 & 0 \\ 
0 & 0 & d_2 & -t_1 & 0 & -m & & \ddots & 0 & 0 \\
\vdots & \vdots & & & \ddots & & \ddots & & t_1 & d_1 \\ 
\vdots & \vdots & & & & \ddots & & \ddots & -d_2 & -t_2\\
0 & 0 & \cdots & \cdots & 0 & 0 & t_2 & -d_1 & m & 0 \\
0 & 0 & \cdots & \cdots & 0 & 0 & d_2 & -t_1 & 0 & -m \\
\end{pmatrix}
\ .
\end{equation}
We thus need to find the eigenvalues $\lambda$ of this matrix
with arbitrary parameters $m,t_1,t_2,d_1,d_2$.

To this end, we write $\mathcal{H}_{\rm BdG} \psi = \lambda \psi$
and we denote the components of $\psi$ by $\psi_{n}$.
The eigenvalue equations give rise to $2L-4$ bulk equations (with $n = 3, \ldots 2L-2$),
\begin{equation}
\begin{cases}
t_2 \psi_{n-2} - d_1 \psi_{n-1} + (m-\lambda) \psi_{n} + t_1 \psi_{n+2} + d_1 \psi_{n+3} = 0 & \text{for $n$ odd}\\
d_2 \psi_{n-3} - t_1 \psi_{n-2} + (-m - \lambda) \psi_{n} - d_2 \psi_{n+1} - t_2 \psi_{n+2} = 0 & \text{for $n$ even} \ .
\end{cases}
\end{equation}
In addition, there are four boundary equations which read, after using the bulk equations (which we will always solve for arbitrary integer $n$)
\begin{align}
- t_2 \psi_{-1} + d_1 \psi_{0} &=0 & - t_1\psi_{2L+1} - d_1 \psi_{2L+2} &= 0 \nonumber\\
- d_2 \psi_{-1} + t_1 \psi_{0} &=0 & + d_2\psi_{2L+1} + t_2 \psi_{2L+2} &= 0 \ .
\end{align}

To obtain the eigenvalues $\lambda$, we use the following concrete ansats for the components of the (right) eigenvectors,
\begin{equation}
\psi^T = \phi^T (x,a) = (x , a x, x^2 , a x^2, \ldots, x^{L}, a x^{L}) \ .
\end{equation}
This ansatz is inspired by the ansatz used in the hermitian case, where $a$ in general is a phase (due to the superconducting nature of the model). Because we deal with the non-hermitian case, $a$ will have arbitrary modulus.
The structure of the powers of $x$ is the standard ansatz used in solving the type of recurrence relations for the bulk of systems with periodicity. In the physics literature, this dates back at least the seminal paper of Lieb, Schultz and Mattis \cite{LiScMa1961}.
Using this ansatz, the `bulk' equations take the following form,
\begin{align}
\label{eq:bulk-gen}
t_2 -a d_1 +(m-\lambda) x + (t_1 + a d_1) x^2 &= 0 &
d_2 - a t_1 + (-m - \lambda) ax + (- d_2 - a t_2) x^2 &= 0 \ .
\end{align}
By eliminating $a$ from these equations, one obtains a fourth order algebraic equation in $x$,
\begin{equation}
\begin{split}
\label{eq:bulk-x}
(d_1 d_2 - t_1 t_2) x^4
- (\lambda (t_1 - t_2) + m (t_1 + t_2)) x^3
+ (\lambda^2 - m^2 - 2 d_1 d_2 - t_1^2 - t_2^2) x^2 \\
+ (\lambda (t_1-t_2) - m (t_1 + t_2)) x 
+ (d_1 d_2 - t_1 t_2) = 0 \ .
\end{split}
\end{equation}
Thus (because the value for $a$ is uniquely determined by a given solution for $x$), we obtain four solutions of the bulk equations, denoted by $(x_i,a_i)$.
The general form of the eigenvector then is $\psi = \sum_{i=1}^4 c_i \phi(x_i,a_i)$. The coefficients $c_i$ can be obtained from the boundary equations,
which now take the form
\begin{align}
(-t_2 + a_1 d_1) c_1 + (-t_2 + a_2 d_1) c_2 + (-t_2 + a_3 d_1) c_3 + (-t_2 + a_4 d_1) c_4 &=0 \nonumber\\
(-d_2 + a_1 t_1) c_1 + (-d_2 + a_2 t_1) c_2 + (-d_2 + a_3 t_1) c_3 + (-d_2 + a_4 t_1) c_4 &=0 \nonumber\\
-(t_1 + a_1 d_1) x_1^{L+1} c_1 - (t_1 + a_2 d_1) x_2^{L+1} c_2 - (t_1 + a_3 d_1) x_3^{L+1} c_3 - (t_1 + a_4 d_1) x_4^{L+1} c_4 &=0 \nonumber\\
(d_2 + a_1 t_2) x_1^{L+1} c_1 + (d_2 + a_2 t_2) x_2^{L+1} c_2 + (d_2 + a_3 t_2) x_3^{L+1} c_3 + (d_2 + a_4 t_2) x_4^{L+1} c_4 &=0 \ .
\end{align}
In practise, one typically does not obtain the explicit values of the $c_i$, but instead uses these equations to determine the eigenvalues.
The pairs $(x_i,a_i)$ implicitly depend on $\lambda$ and the parameters in the model.
The condition that the boundary equations have a non-trivial solution for the $c_i$ then turns into
an equation for the possible eigenvalues $\lambda$. In particular, we write the equations as
$M c = 0$, where
$c^T = (c_1,c_2,c_3,c_4)$ and
\begin{equation}
\label{eq:m-matrix}
M = \begin{pmatrix}
-t_2 + a_1 d_ 1 & -t_2 + a_2 d_ 1 & -t_2 + a_3 d_ 1 & -t_2 + a_4 d_ 1 \\
-d_2 + a_1 t_1 & -d_2 + a_2 t_1 & -d_2 + a_3 t_1 & -d_2 + a_4 t_1 \\
-(t_1 + a_1 d_1) x_1^{L+1} & -(t_1 + a_2 d_1) x_2^{L+1} & -(t_1 + a_3 d_1) x_3^{L+1} & -(t_1 + a_4 d_1) x_4^{L+1} \\
(d_2 + a_1 t_2) x_1^{L+1} & (d_2 + a_2 t_2) x_2^{L+1} & (d_2 + a_3 t_2) x_3^{L+1} & (d_2 + a_4 t_2) x_4^{L+1}
\end{pmatrix} \ .
\end{equation}
The condition to have a non-trivial solution for the $c_i$ is $\det (M) = 0$, which determines $\lambda$ via
the parameters $(x_i,a_i)$.
In the sections below, we perform the analysis to characterise $\lambda$ for various different cases.

Before doing so, we mention, for later use, that the eigenvalues $\lambda$ of the periodic version of the model take a simple
form in terms of the momentum $k$, namely
\begin{equation}
\label{eq:lambda-periodic}
\lambda_\pm (k) =  i (t_1-t_2) \sin(k) \pm \sqrt{4 d_1 d_2 \sin(k)^2 + (m + (t_1+t_2)\cos(k) )^2} \ .
\end{equation}
These eigenvalues corresponding to the model with periodic boundary conditions are obtained by first performing a (discrete)
Fourier transform. This results in the following $2\times 2$ matrix
\begin{equation}
\begin{pmatrix}
m + t_1 e^{i k} + t_2 e^{-i k} & d_1 (e^{i k} + e^{-ik}) \\
-d_2 (e^{i k} + e^{-ik}) & -m - t_1 e^{-i k} - t_2 e^{i k} 
\end{pmatrix}
\ .
\end{equation}
The eigenvalues of this matrix are indeed given by Eq.~\eqref{eq:lambda-periodic}.

\section{Solution for finite $L$, with $t_2=t_1$ or $d_1  d_2 = 0$}
\label{sec:finite-size}

In this section, we start by presenting the full solution of the model, for an open chain of size $L$, but with the restriction that
$t_2 = t_1$. In this section, we use the notation $t = t_1 = t_2$, to remind the reader of this restriction.
We use the method of Lieb, Schultz and Mattis \cite{LiScMa1961}, which was used to study the hermitian Kitaev chain with
longer range hopping in \cite{MaAr2018}. We use the approach taken in the latter paper, or rather, repeat the calculation.
We only need to note that the restriction used in that paper, namely that the chemical potential and the hopping parameter are real,
can in fact be dropped, without invalidating the solution. Thus, we allow complex parameters $m, t=t_1 = t_2, d_1, d_2$.
We note these parameters {\em do not} include all the hermitian cases, the hermitian case with complex hopping parameters
is excluded. However, the solution {\em does} include non-hermitian cases.

Thus, the goal of this section is to describe the eigenvalues of the matrix $\mathcal{H}_{\rm BdG}$ of Eq.~\eqref{eq:nh-top-sc-matrix},
with (possibly complex) parameters $m, t = t_1 = t_2, d_1, d_2$. Here, we are very brief, and refer to \cite{MaAr2018} for the details.

We write the ansatz for the eigenvalues as $\lambda_\alpha = \pm \sqrt{4 d_1 d_2 \sin^2(\alpha) + (m+ 2 t \cos(\alpha))^2}$,
inspired by the solution for the periodic case with $t = t_1 = t_2$, Eq.~\eqref{eq:lambda-periodic}.
To find the values of $\alpha$, we have to solve the `bulk' equation Eq.~\eqref{eq:bulk-x}, for the wave functions.
The `boundary' equations then give two equations that determine $\alpha$.
The bulk equation actually has four solutions, $e^{i \alpha}, \ e^{-i \alpha}, \ e^{i \beta}, \ e^{-i \beta}$, where the values of
$\alpha$ and $\beta$ are related by the following equation
\begin{equation}
\label{eq:coseqn}
2 \cos (\alpha) + 2 \cos(\beta) = \frac{2mt}{d_1 d_2 - t^2} \ .
\end{equation}
Often, we introduce the notation $x = e^{i \alpha}$ and $y = e^{i \beta}$, because in this way, the equations for
$x$ and $y$ are polynomial equations, while those for $\alpha$ and $\beta$ are trigonometric.

The boundary equations lead to a determinant that should be zero, $\det(M) = 0$ with $M$ given by Eq.~\eqref{eq:m-matrix},
giving the second equation that we need to determine
$\alpha$ and $\beta$ (or $x$ and $y$). To describe this equation, we introduce the following sine ratios
\begin{equation}
{\rm sr} (L,\alpha) = \frac{\sin(L \alpha)}{\sin(\alpha)} = x^{-L+1} + x^{-L+3} + \cdots + x^{L-3} + x^{L-1} \ ,
\end{equation} 
and similar for $\beta$ and $y$.
Using these functions, the determinant equation takes the following form
\begin{equation}
\label{eq:deteqn}
(d_1 d_2 - t^2) {\rm sr} (L+1,\alpha) {\rm sr} (L+1,\beta) +
4 d_1 d_2 \sum_{j=1}^{L} (L+1-j) {\rm sr} (j,\alpha) {\rm sr} (j,\beta) = 0 \ .
\end{equation}
We note that the equivalent equation in \cite{MaAr2018} takes a different (and more complicated) form, because here, we
simplified the boundary equations before forming the determinant equation that finally gives the solutions.

To determine the eigenvalues $\lambda_{\alpha}$,
we need to solve Eqs.~\eqref{eq:coseqn} and \eqref{eq:deteqn} simultaneously. 
These equations are (separately) invariant under $x \leftrightarrow y$, $x \leftrightarrow 1/x$ and $y \leftrightarrow 1/y$.
So, the $8L$ solutions of these equations correspond to $L$ plus/minus eigenvalue pairs, so $2L$ eigenvalues
as needed.

We note that the functions ${\rm sr} (L,\alpha)$ are, by definition, closely related to the
Chebyshev polynomials of the second kind $U_n$, namely ${\rm sr} (L,\alpha) = U_L (\cos \alpha)$ \cite{Ch1853}.
Also, the sum over $j$ in Eq.~\eqref{eq:deteqn} `can be done', leading to the following result
\begin{align}
& (d_1 d_2 - t^2) {\rm sr} (L+1,\alpha) {\rm sr} (L+1,\beta) + \\
&+\frac{d_1 d_2}{(\cos\alpha-\cos\beta)^2}
\bigl(
2 - 2 {\rm sr}(L+1,\alpha){\rm sr}(L+1,\beta) + {\rm sr}(L+2,\alpha){\rm sr}(L,\beta) + {\rm sr}(L,\alpha){\rm sr}(L+2,\beta) 
\bigr) = 0 \ .\nonumber
\end{align}

Before continuing the analysis of the model for infinite system size, we briefly consider the case with either $d_1 = 0$ or
$d_2 = 0$, but not necessarily both, and otherwise arbitrary parameters (so we allow $t_1 \neq t_2$ here).
In this case, the model is closely related to the Hatano-Nelson model \cite{HaNe1996,HaNe1997,HaNe1998}, for which an exact solution
that interpolates between open and periodic boundary conditions
was presented in \cite{EdAr2022}.
Using the techniques of that paper, we find that for $d_2 = 0$, the eigenvalues do not depend on $d_1$ (and the other way around),
but a subset of the eigenvectors does depend on $d_1$. The eigenvalues are given by
\begin{equation}
\label{eq:d1d2-zero}
\lambda_{\pm,j} = \pm m + 2 \sqrt{t_1}\sqrt{t_2} \cos \bigl( \frac{j \pi}{L+1}\bigr) \ ,
\end{equation}
for $j = 1, 2, \ldots, L$. We already note that for $d_1 = 0$ or $d_2 = 0$, the system does not have an isolated zero mode, and the
eigenvectors show skin-effect when $| t_1 | \neq | t_2 |$. Finally, we note that in the case $d_1 d_2 = 0$, we can easily take the
limit $L\rightarrow\infty$. That is, the eigenvalues in the infinite size limit form two line segments in the complex plane.

\section{Condition for the presence of zero modes.}
\label{sec:zero-modes}

In this section, we determine under which conditions the model has a zero mode.
Just as in the hermitian Kitaev chain \cite{Ki2001}, the zero modes exhibit an exponential decay
away from the boundaries, and correspond to topological edge states. 

We consider arbitrary complex parameters, and work directly in the
$L\rightarrow\infty$ limit. In finite systems, the energy of a zero mode
decays exponentially with system size, but because we work in the limit $L\rightarrow\infty$,
we put $\lambda = 0$ identically.
For $\lambda = 0$, it turns out that one can obtain the solutions
of the bulk equations, Eq.~\eqref{eq:bulk-gen}, explicitly 
\begin{align}
x_{-,-} &= \frac{- m - \sqrt{4 d_1 d_2 + m^2 - 4 t_1 t_2}}{(t_1 + t_2) - \sqrt{4 d_1 d_2 + (t_1 - t_2)^2}}
&
a_{-} &= \frac{-(t_1-t_2) - \sqrt{4 d_1 d_2 + (t_1 - t_2)^2}}{2d_1}
\\
x_{+,-} &= \frac{- m + \sqrt{4 d_1 d_2 + m^2 - 4 t_1 t_2}}{(t_1 + t_2) - \sqrt{4 d_1 d_2 + (t_1 - t_2)^2}}
&
a_{-} &= \frac{-(t_1-t_2) - \sqrt{4 d_1 d_2 + (t_1 - t_2)^2}}{2d_1}
\\
x_{-,+} &= \frac{- m - \sqrt{4 d_1 d_2 + m^2 - 4 t_1 t_2}}{(t_1 + t_2) + \sqrt{4 d_1 d_2 + (t_1 - t_2)^2}}
&
a_{+} &= \frac{-(t_1-t_2) + \sqrt{4 d_1 d_2 + (t_1 - t_2)^2}}{2d_1}
\\
x_{+,+} &= \frac{- m + \sqrt{4 d_1 d_2 + m^2 - 4 t_1 t_2}}{(t_1 + t_2) + \sqrt{4 d_1 d_2 + (t_1 - t_2)^2}}
&
a_{+} &= \frac{-(t_1-t_2) + \sqrt{4 d_1 d_2 + (t_1 - t_2)^2}}{2d_1}
\ ,
\end{align}
where $x_{+,+} = 1/x_{-,-}$ and $x_{-,+} = 1/x_{+,-}$.

We can now take linear combinations of these solutions, to satisfy the boundary equations.
We solve the left boundary equations exactly, and demand that $|x| < 1$, so that the right boundary equations are satisfied in the thermodynamic limit.

We find the following two solutions
\begin{align}
\Psi^T_{-} &= \phi^T (x_{-,-},a_{-}) - \phi^T (x_{+,-},a_{-})
\\
\Psi^T_{+} &= \phi^T (x_{-,+},a_{+}) - \phi^T (x_{+,+},a_{+})
\ .
\end{align}

In order to satisfy the boundary equations on the right hand side, we need that either $|x_{-,-}| < 1$ and $|x_{+,-}| < 1$ such that
$\Psi^T_{-}$ is a zero mode or that 
$|x_{-,+}| < 1$ and $|x_{+,+}| < 1$ such that
$\Psi^T_{+}$ is a zero mode.
This leads to constraints on the parameters in the model.

\subsection{The hermitian case}
To analyse under which conditions there is a zero mode, we first consider the hermitian case, which is simpler.
That is, we assume that $m\in \mathbb{R}$.
We also write $t_1 = t e^{i \phi_t}$, $t_2 = t e^{-i \phi_t}$, $d_1 = d e^{i \phi_d}$ and $d_2 = d e^{-i \phi_d}$,
with $t\geq 0,d\geq 0,0\leq\phi_t<2\pi,0\leq\phi_d<2\pi$ all real. We then have
\begin{align}
x_{-,-} &= \frac{- m/2 - \sqrt{d^2 + (m/2)^2 - t^2}}{t \cos\phi_t - \sqrt{d^2 - t^2 \sin^2\phi_t}}
&
x_{+,-} &= \frac{- m/2 + \sqrt{d^2 + (m/2)^2 - t^2}}{t \cos\phi_t - \sqrt{d^2 - t^2 \sin^2\phi_t}}
\\
x_{-,+} &= \frac{- m/2 - \sqrt{d^2 + (m/2)^2 - t^2}}{t \cos\phi_t + \sqrt{d^2 - t^2 \sin^2\phi_t}}
&
x_{+,+} &= \frac{- m/2 + \sqrt{d^2 + (m/2)^2 - t^2}}{t \cos\phi_t + \sqrt{d^2 - t^2 \sin^2\phi_t}} \ .
\end{align}
Analysing the conditions that imply the existence of a zero mode (that is, either 
$|x_{-,-}| < 1$ and $|x_{+,-}| < 1$, or $|x_{-,+}| < 1$ and $|x_{+,+}| < 1$), we consider four
different cases, determined by the expressions under the square roots being positive or negative.
We find that it is necessary to have $d^2 > t^2 \sin^2\phi_t$, while $d^2 + (m/2)^2 - t^2$
can have either sign. In addition, it is necessary that $m^2 < 4t^2 \cos^2 \phi_t$.
Under these conditions, we have that $| x_{-,+} | < 1$ and $| x_{+,+} | < 1$ if $\cos\phi_t > 0$.
If on the other hand $\cos\phi_t < 0$, we have that $| x_{-,-} | < 1$ and $| x_{+,-} | < 1$ instead.
From this, we also find that it is necessary to have $\cos\phi_t \neq 0$.
This condition is, however, already implied by $m^2 < 4t^2 \cos^2 \phi_t$.
Summarising, we find the following conditions in order that the system has a zero mode
in the hermitian case
\begin{align}
m^2 &< 4 t^2 \cos^2\phi_t & d^2 & > t^2 \sin^2\phi_t \ .
\end{align}
In the case of real hopping parameters, $\phi_t = 0, \pi$, this reduces to the well known
conditions $|m| < 2 |t|$ and $|d| > 0$ \cite{Ki2001}.

\subsection{The general case}
In the general case we have $x_{\pm,\pm} = \frac{- m \pm \sqrt{4 d_1 d_2 + m^2 - 4 t_1 t_2}}{(t_1 + t_2) \pm \sqrt{4 d_1 d_2 + (t_1 - t_2)^2}}$.
To simplify the expressions (here and below), we introduce $D_2 = (d_1 d_2 - t_1 t_2)$, and put a factor $1/\sqrt{-4D_2}$
(which is in general complex) under the square roots.
This of course might introduce an additional sign for the square root terms,
but because we need simultaneous conditions for the different signs,
this does influence the range of parameters for which there is a zero mode
(however, the values of the various $x_{\pm,\pm}$ might be swapped).
With these caveats, we write
\begin{align}
x_{\pm,\pm} &= \frac{y_1 \pm \sqrt{y_1^2 - 1}}{y_2 \pm \sqrt{y_2^2 - 1}}
&
y_1 &= -\frac{m}{\sqrt{-4 D_2}}
&
y_2 &= \frac{t_1+t_2}{\sqrt{-4 D_2}}
\ .
\end{align}

We focus on the expression $y \pm \sqrt{y^2-1}$.
Modulo the values at the branch cuts, we can write
\begin{equation}
y \pm \sqrt{y^2-1} =
\begin{cases}
e^{\pm i \arccos(y)} & \Re(y) \Im(y) > 0 \\
e^{\mp i \arccos(y)} & \Re(y) \Im(y) < 0 \ .
\end{cases}
\end{equation}
Because we are interested in the absolute values of $x_{\pm,\pm}$, we have to investigate $\Im (\arccos(y_1))$ and $\Im (\arccos(y_2))$.
We demand that either $|x_{-,-}| < 1 \wedge |x_{+,-}| < 1$ or $|x_{-,+}| < 1 \wedge |x_{+,+}| < 1$.
Analysing these conditions, one finds that there is a zero mode when
$|\Im (\arccos(y_1))|  < |\Im (\arccos(y_2))|$,
or in terms of the parameters of the model,
\begin{equation}
\label{eq:zero-mode-criterion}
\Bigl|\Im \Bigl(\arccos\bigl(-\frac{m}{2\sqrt{t_1 t_2 - d_1 d_2}}\bigr)\Bigr)\Bigr| 
<
\Bigl|\Im \Bigl(\arccos\bigl(\frac{t_1 + t_2}{2\sqrt{t_1 t_2 - d_1 d_2}}\bigr)\Bigr)\Bigr| \ .
\end{equation}

We close this section by commenting on the case $d_1 d_2 = 0$. The criterion Eq.~\eqref{eq:zero-mode-criterion}
is not in any way singular when $d_1 d_2 = 0$. However, we know that in this case, the model is closely related
to the Hatano-Nelson model \cite{HaNe1996,HaNe1997,HaNe1998}, see Sec.~\ref{sec:finite-size}.
In particular, the eigenvalues for the finite size system
are given by Eq.~\eqref{eq:d1d2-zero}, showing that for generic parameters, the model does not have a zero mode
for chains of finite size. However, Eq.~\eqref{eq:zero-mode-criterion} can certainly be satisfied when $d_1 d_2 = 0$.
This means that the zero mode only occurs in the limit $L \rightarrow \infty$. We verified this behaviour in the
following way.

We first picked parameters, with $d_1 \neq 0$ and $d_2 = 0$, such that Eq.~\eqref{eq:zero-mode-criterion} is
satisfied (i.e., there is a zero mode when $L \rightarrow \infty$). For a large, but finite system, the eigenvalues are given
by Eq.~\eqref{eq:d1d2-zero}, and generically, there are no eigenvalues that tend to zero when increasing the system size
(which would be the case for a zero mode that is present already for finite system size). However, upon changing $d_2$ to
a value such that $|d_2|$ is much smaller than the absolute values of all the other parameters in the system, the spectrum
reorganises itself in such a way that there is a zero mode that is present at finite system size. That is, there is a pair of
eigenvalues that tends to zero upon increasing the system size.

In contrast to this, if we pick parameters with $d_1 \neq 0$ and $d_2 = 0$, such that Eq.~\eqref{eq:zero-mode-criterion} is
not satisfied (i.e., there is no zero mode when $L \rightarrow \infty$), the spectrum only changes slightly, when slightly changing
$d_2$ away from zero. In particular, no eigenvalues appear that tend to zero upon increasing the system size, in agreement with
the absence of a zero mode.

\section{Solution in the $L\rightarrow\infty$ limit}
\label{sec:Linfinity}

In section \ref{sec:finite-size}, we obtained a compact characterisation of the eigenvalues
for chains of arbitrary finite length, under the restriction that $t_1 = t_2$, but otherwise
arbitrary complex parameters. Because obtaining the eigenvalues for large
non-hermitian systems is often numerically unstable, we focus in this section on
obtaining a compact characterisation for the eigenvalues in the $L\rightarrow\infty$ limit.
We do this for arbitrary complex parameters, so in this section we relax the constraint
$t_1 = t_2$ that we imposed above.
The result of this section is a complete characterisation of the eigenvalues in terms of
three polynomial equations, which can be solved numerically straightforwardly, without
stability issues.

The starting point is the condition $\det(M) = 0$, where $M$ is given by Eq.~\eqref{eq:m-matrix}.
Evaluating the determinant gives (after dropping a factor $(d_1 d_2 - t_1 t_2)^2$)
\begin{equation}
\begin{split}
 (a_1- a_2)(a_3 - a_4) (x_1^{L+1} x_2^{L+1}+ x_3^{L+1} x_4^{L+1}) \\ 
- (a_1 - a_3)(a_2 - a_4) (x_1^{L+1} x_3^{L+1}+ x_2^{L+1} x_4^{L+1}) \\
+ (a_1 - a_4)(a_2 - a_3) (x_1^{L+1} x_4^{L+1} + x_2^{L+1} x_3^{L+1}) = 0 \ .
\end{split}
\label{eq:det-condition}
\end{equation}

We recall that the $(x_i,a_i)$ are determined by the bulk equations~\eqref{eq:bulk-gen}.
To analyse this condition, we order the $x_i$ according to their absolute values, $|x_1| \geq |x_2| \geq |x_3| \geq |x_4|$.
From the bulk equation~\eqref{eq:bulk-x}, it follows that $x_1 x_2 x_3 x_4 = 1$.

Generically, the condition Eq.~\eqref{eq:det-condition} is dominated by $x_1^{L+1} x_2^{L+1}$.
This means that we obtain the condition $(a_1- a_2)(a_3 - a_4) = 0$.
We find that, for generic parameters in the model, the only way in which one can have a double root for $a$ is when
$\lambda = 0$, i.e., in the case of a zero mode. 

To show this, we need to analyse under which conditions there is a solution for which two values of $a$ are identical,
and check that they correspond to the solutions for $x$ which are largest in absolute value.
One can eliminate $x$ from the bulk equations Eqs.~\eqref{eq:bulk-gen}, giving rise to a fourth order polynomial in $a$,
which we denote as $p_4(a)$. To find a double zero, one needs that $p_4(a)$ and $\frac{d p_4(a)}{da}$ have a common zero.
This in turn occurs when the {\em resultant} of $p_4(a)$ and $\frac{d p_4(a)}{da}$ is zero,
${\rm Res} (p_4(a),\frac{d p_4(a)}{da}) = 0$.

One can write ${\rm Res} (p_4(a),\frac{d p_4(a)}{da}) = 16 d_1^4 d_2^2 \lambda^4 (t_1 + t_2)^4 (\lambda+m -t_1-t_2)(\lambda+m + t_1+t_2) f_8(\lambda)$,
where $f_8(\lambda)$ is an eighth order polynomial in $\lambda$, which also depends on all the parameters in the model
(the precise form of this polynomial is complicated, and not interesting for our purposes).
We find that two values of $a$ coincide for $d_1 \, d_2 = 0$, for $\lambda = 0$, for $t_2 = - t_1$ 
as well as for ten special values of $\lambda$ that depend on the parameters of the model.
These ten special values include $\lambda = -m \pm (t_1 + t_2)$. We do not consider these ten special values,
because we are interested the generic eigenvalues of the model.

The case $d_1 \, d_2 = 0$ was treated in Sec.~\ref{sec:finite-size}.
The case $\lambda = 0$, i.e., the zero modes, was analysed in detail in Sec.~\ref{sec:zero-modes} above.
In particular, we obtained a condition for the parameters in the model,
such that there is an actual zero mode. This condition corresponded to the conditions $|x_1 | > 1$ and $|x_2| > 1$.
When there is no zero mode, we find the double solution for $a$ actually corresponds to $a_3 = a_1$ or $a_4 = a_1$, so that
Eq.~\eqref{eq:det-condition} is not satisfied in the thermodynamic limit, despite the fact that there is a double solution for $a$.

When analysing the case $t_2 = -t_1$, we come to the same conclusion,
Eq.~\eqref{eq:det-condition} is not satisfied in the thermodynamic limit, despite the fact that there is a double solution for $a$.

One is left to wonder how one can satisfy Eq.~\eqref{eq:det-condition} when $\lambda \neq 0$ for arbitrary parameters?
The answer is that we made an implicit assumption, namely we assumed that $|x_2| > |x_3|$,
from which it followed that $x_1^{L+1} x_2^{L+1}$ dominates the expression. Under the condition
that $|x_2| = |x_3|$, the condition Eq.~\eqref{eq:det-condition} is dominated by more terms,
such that solutions can be found for which $(a_1- a_2)(a_3 - a_4) \neq 0$.

This observation allows us to obtain a rather compact representation of the eigenvalues $\lambda$
in the limit of large system size, $L\rightarrow \infty$. Namely, we demand that two of the roots
$x_i$ of the bulk equation Eq.~\eqref{eq:bulk-x} have the same absolute value.

That is, we write
\begin{align}
\label{eq:roots-infinite-system}
x_1 &= \frac{s}{\kappa} &
x_2 &= \kappa e^{i \alpha} &
x_3 &= \kappa e^{-i \alpha} &
x_4 &= \frac{1}{s \kappa}  \ ,
\end{align}
where $\kappa$ and $s$ are in general complex, while $0 \leq \alpha < 2 \pi$ is real.
Clearly $x_1 x_2 x_3 x_4 = 1$,
as required by the bulk Eq.~\eqref{eq:bulk-x}.

The condition $x_1 x_2 x_3 x_4 = 1$ is one of the Vieta equations,
relating the roots of a polynomial to its coefficients \cite{Vi1646}.
Before continuing our analysis, we quickly explain the Vieta equations.
For ease of presentation, we do this for a fourth order polynomial, as the generalisation to
the arbitrary case is clear. We write the polynomial in terms of its coefficients $a_i$ and its
roots $r_j$ as follows
\begin{equation}
P(x) = a_4 x^4 + a_3 x^3 + a_2 x^2 + a_1 x + a_0 = a_4 \prod_{j=1}^{4} (x-r_j) \ .
\end{equation}
Expanding the right hand side, and comparing with the coefficients leads to the following equations
\begin{align}
r_1 + r_2 + r_3 + r_4 &= -a_3/a_4 \\
r_1 r_2 + r_1 r_3 + r_1 r_4 + r_2 r_3 + r_2 r_4 + r_3 r_4 &= a_2/a_4 \\
r_1 r_2 r_3 + r_1 r_2 r_4 + r_1 r_3 r_4 + r_2 r_3 r_4 &= -a_1/a_4 \\
r_1 r_2 r_3 r_4 &= a_0/a_4 \ .
\end{align}
These are the Vieta equations, which relate the coefficients of a polynomial to its roots.
We note that the left hand sides correspond to the elementary symmetric polynomials evaluated at the
roots of the original polynomial $P(x)$.

In the case at hand we have $a_0 = a_4$, leading to the fourth Vieta equation $x_1 x_2 x_3 x_4 = 1$ as stated above.
The remaning three Vieta equations for the roots can then be written (after taking a linear combination) as
\begin{align}
&\bigl( \kappa + \frac{1}{\kappa}\bigr) \bigl[ (s + \frac{1}{s}) + 2 \cos\alpha \bigr] = \frac{2 m (t_1 + t_2)}{d_1 d_2 - t_1 t_2} \nonumber\\
&\bigl( \kappa - \frac{1}{\kappa}\bigr) \bigl[ (s + \frac{1}{s}) - 2 \cos\alpha \bigr] = \frac{2 \lambda (t_1 - t_2)}{d_1 d_2 - t_1 t_2} \nonumber\\
&( \kappa + \frac{1}{\kappa})^2 + 2 \cos\alpha (s + \frac{1}{s}) = \frac{\lambda^2 - m^2 - (t_1 + t_2)^2}{d_1 d_2 - t_1 t_2} \ .
\label{eq:vieta}
\end{align}
The curve(s) in the complex plane determined by the eigenvalues $\lambda$ in the limit $L \rightarrow\infty$ are
obtained by varying $0 \leq \alpha < 2\pi$ (we note that $\alpha$ plays the role of the momentum $k$ in the periodic case).
In principle, one can eliminate $\kappa$ and $s$ from the Vieta equations, to obtain an equation for
$\lambda$ in terms of $\alpha$ and the parameters of the model. This results in a forth order equation in $\lambda^2$,
which is not insightful.
In practise, if one wants to obtain the actual curve, one can simply solve the Eqs.~\eqref{eq:vieta} numerically.
It should be noted that not all solutions for $\lambda$ correspond to actual eigenvalues. 
This is because it needs to be
checked that the roots are ordered as $|x_1| \geq |x_2| = |x_3| \geq |x_4|$. There are two branches for which
$|x_2| = |x_3| \geq |x_1|, |x_4|$ or $|x_1|, |x_4| \geq |x_2| = |x_3| $,
which {\em do not} lead to actual eigenvalues.
Thus, we need the solutions of the Vieta Eqs.~\eqref{eq:vieta} such that $1/|s| \leq |\kappa|^2 \leq |s|$. Equivalently, one
can instead require that $|s| \leq |\kappa|^2 \leq 1/|s|$.
Despite the fact that there are two non-physical branches, that do not lead to actual eigenvalues of the model,
these non-physical branches play an important role in explaining the geometry of the actual curves the eigenvalues lie on.
Without the additional constraint that the roots need to satisfy $|x_1| \geq |x_2| = |x_3| \geq |x_4|$, one would obtain
smooth curves for the eigenvalues in the complex plane. With the additional necessary constraint however,
one finds that the actual curves on which the eigenvalues lie exhibit branching points. This is because these actual
curves `jump from one branch to another'.

\begin{figure}[t]
\includegraphics[width=0.8\textwidth]{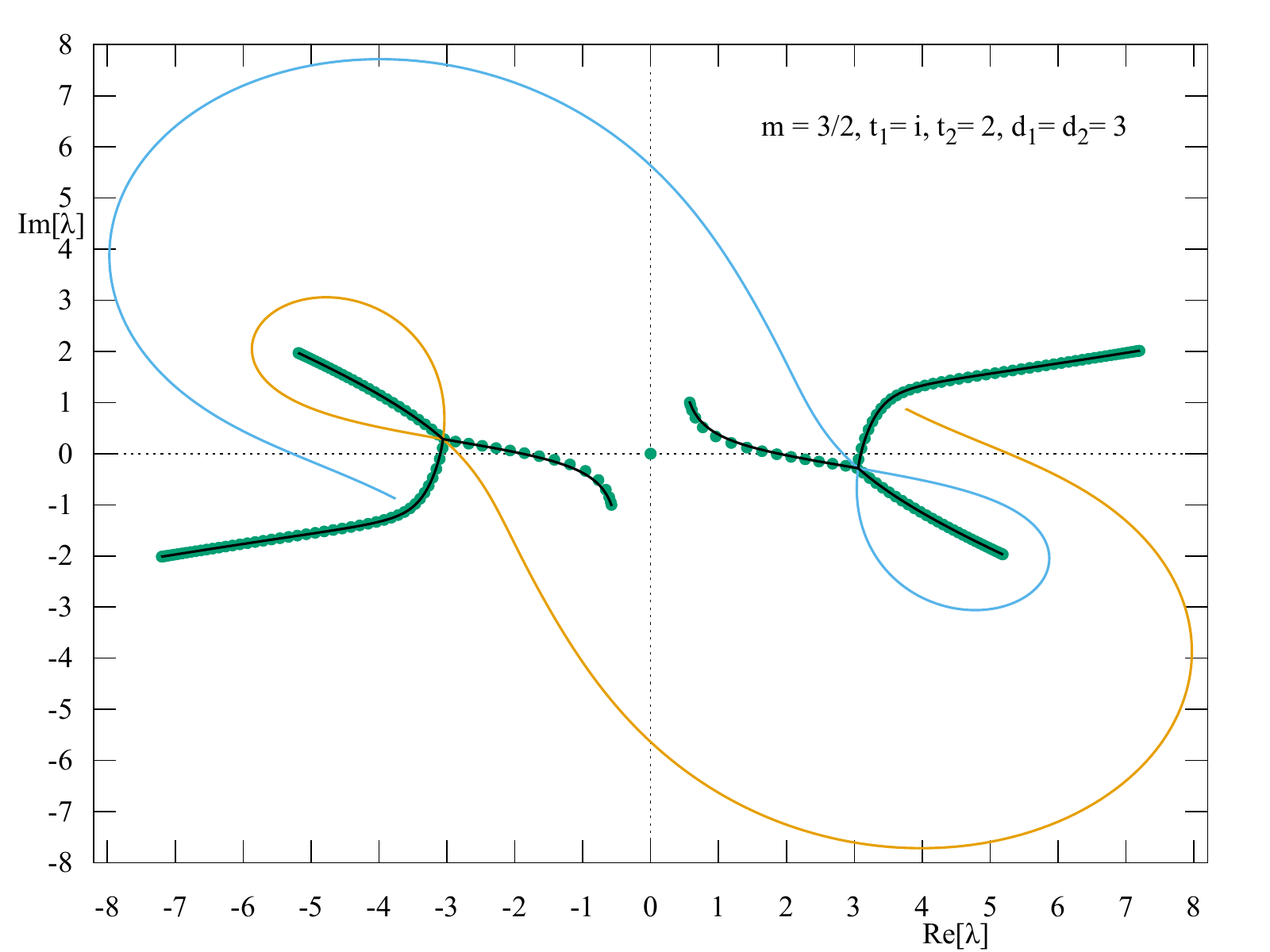}
\caption{Plot for the non-hermitian Kitaev Chain with parameters $m = 3/2$, $t_1 = i$, $t_2 = 2$, $d_1 = d_2 = 3$.
The black lines correspond to the eigenvalues of the infinite system. The blue and yellow lines correspond to
solutions for $\lambda$ of the Vieta equations, that do not correspond to actual eigenvalues of the model (as explained
in the main text). The green dots correspond to eigenvalues of the finite system with $L = 100$.
}
\label{fig:infinite-size-example}
\end{figure}

To illustrate this, we plot the solutions for $\lambda$ of the Vieta Eqs.~\eqref{eq:vieta} for the parameters
$m = 3/2$, $t_1 = i$, $t_2 = 2$ and $d_1 = d_2 = 3$ in Fig.~\ref{fig:infinite-size-example}. To generate the plot,
we vary $\alpha$ over $0 \leq \alpha < 2\pi$ in steps of $2\pi/1000$. Using the solutions for $\kappa$ and $s$,
we determine in which way the absolute values of $x_i$ are ordered. The black lines correspond to the ordering
$|x_1| \geq |x_2| = |x_3| \geq |x_4|$, that is, to actual eigenvalues of the model. The blue and yellow lines correspond
to the other two orderings, that do not lead to eigenvalues of the model. Finally, the green dots correspond to the
eigenvalues of the finite chain with $L=100$. Fig.~\ref{fig:infinite-size-example} clearly shows that the green dots
closely follow the black lines, with only small deviations, caused by finite size effects. In addition, it is clear that the
blue and yellow lines do not correspond to actual eigenvalues of the model. We also note the presence of the zero mode
in the spectrum of the finite model. Zero modes do not correspond to solutions of the Vieta equations, as explained above.

The fact that the solutions for $x$ come in three different `branches', with only one corresponding to actual eigenvalues
of the model, nicely explains the presence of the `branching points' that are present in the spectra of the non-hermitian
Kitaev chain. We show a more intricate example of this in Sec.~\ref{sec:summary} below. There, we also discuss the
known stability problems of finding the eigenvalues of a large finite (non-hermitian) system. 

We close this section by noting that if we know the spectrum for a given set of parameters, we in fact know the spectrum
for a one-parameter set of parameters. In particular, the right hand sides of the Vieta Eqs.~\eqref{eq:vieta} are invariant
under
\begin{align}
\label{eq:paramters-phase-rotation}
m &\rightarrow m e^{i \phi} & t_s &\rightarrow t_s e^{i \phi} & t_d &\rightarrow t_d e^{i \phi} & d_1 d_2 &\rightarrow d_1 d_2 e^{2i \phi}
& \lambda &\rightarrow \lambda e^{i \phi} \ ,
\end{align}
meaning that if one changes phases of the parameters in the way indicated, the whole spectrum is rigidly phase rotated.
We note that this basically corresponds to multiplying the whole matrix Eq.~\eqref{eq:nh-top-sc-matrix} by a constant phase,
which leaves the eigenvectors invariant (one can of course also rescale the spectrum in the same way).

\section{Presence of the skin effect}
\label{sec:skin-effect}

Because we have a rather compact characterisation of the eigenvalues of the model in the thermodynamic limit
we can, in principle, analyse in detail for which parameters the model has a skin effect.
Eigenvectors that exhibit skin effect are exponentially localised to the boundary of the systems.
We determine the parameters for which the system exhibits a skin effect, and if so, to which eigenvectors
this pertains (there can be cases where only a subset of eigenvectors exhibit skin effect).

In solving the model in the infinite size limit, we obtained that two roots of the bulk equation are equal in absolute
value, see Eq.~\eqref{eq:roots-infinite-system}.
Because the roots $x_2$ and $x_3$ with $|x_2| = |x_3|$ determine the corresponding eigenvector, we find that there is no 
skin effect when $|x_2| = |x_3| = 1$. The eigenvalues of the eigenstates that do not show a skin effect, lie on the curves
of the eigenvalues in the periodic case $\lambda_\pm (k)$, as given by Eq.~\eqref{eq:lambda-periodic}.
We are interested in determining the parameters of the model, for which there are eigenstates that do not show the skin
effect for extended ranges of $k$. We will not in general try to locate isolated points.

There is a long history of determining the location of the roots of polynomials in the complex plane.
One of the main reasons for this, is that it is used extensively in system analysis, in particular for systems that
are linear and time invariant. Such a system is stable, if the output is bounded, even in the limit when the evolved
time goes to infinity. This is the case when all the roots of the characteristic polynomial corresponding to the system
have negative real parts. 
Several algorithms exist to determine, how many roots have a negative real part, {\em without} having to determine
the actual roots. Routh and Hurwitz independently developed such algorithms, leading to the Routh–Hurwitz stability criterion \cite{Ro1877,Hu1895},  which determines if all the roots of a polynomial have negative real parts. One way to derive the criterion
is to construct the sequence of Sturm polynomials associated with the polynomial under investigation \cite{St1829}.

The  algorithm we focus on is tailored to determine the number of roots within, on and outside of the unit circle.
This is achieved by using a conformal map, that maps the imaginary axis to the unit circle.
In particular, the algorithm we use is due to Bistritz \cite{Bi2002}, but also in the case of finding the number
of roots inside the unit circle, the topic has a long history, dating back a century at least \cite{Co1922}.

For a polynomial with explicit coefficients, the algorithm fully determines the number of roots of each
`type'. We however, would like to determine the number of roots on the unit circle as a function of the parameters
in the model. This is a harder problem, and although we believe we determined all cases for which there are at least
two roots on the unit circle, we do not have a proof for this in the general case with complex parameters.

We do not repeat the full Bistritz algorithm here (which can be found in \cite{Bi2002}), because it is a bit lengthy, and
we do not need most of the details for our purposes. Therefor, we focus instead on the parts of the
algorithm that are relevant for our analysis. 

The starting point of the algorithm is a polynomial of degree $n$, $P_n(x)$, and we assume that $P_n(1) = 1$. As long as
$P_n(1) \neq 0$, we can always rescale $P_n(x)$ as necessary. If $P_n(1) = 0$, we can factorise out the root $x=1$.
From the polynomial $P_n(x)$, a set of polynomials $T_i (x)$, of degree $i$ with $i = n, n-1,\ldots, 0$ is constructed.
In this paper, we do not describe the actual algorithm to determine the polynomials $T_i (x)$, but simply state the
results and refer to \cite{Bi2002} for the details. 

In general, the algorithm to determine the polynomials $T_i(x)$ can be `regular' or `singular'.
For now, we assume that the algorithm is `regular' and discuss the singular case below.
We assume that we obtained the polynomials $T_i(x)$ explicitly.

From the polynomials $T_i (x)$ one forms the sequence
\begin{equation}
A = \{ T_n (1), T_{n-1} (1), \ldots T_1 (1), T_0(1) \} \ .
\end{equation}
The algorithm to define the $T_i (x)$ guarantees that $T_i(1)$ is real for all $i$.
Therefore, we can define $\nu_n$ as the number of sign changes in the sequence $A$.
The number of zeroes of $P(x)$ inside the unit circle is then given by $\alpha_n = n - \nu_n$,
while the number of zeroes of $P(x)$ outside the unit circle is given by $\gamma_n = \nu_n$.

If the polynomial $P_n(x)$ has one or more roots on the unit circle, the algorithm is `singular'.
In particular, the algorithm is singular at level $s$ if $T_s (0) \neq 0$ and $T_{s-1} (x) \equiv 0$.
In this case, the algorithm proceeds in a slightly different manner, and one obtains different
polynomials $T'_{s-1} (x), T'_{s-2} (x), \ldots T'_{1} (x), T'_{0} (x)$ (which also have the property
that $T'_{i} (1)$ is real for all $i$).
In this case, one defines
\begin{align}
A &= \{ T_n (1), T_{n-1} (1), \ldots, T_{s}(1), T'_{s-1}(1), T'_{s-2}(1) \ldots T'_1 (1), T'_0(1) \} \\
B &= \{ T_{s}(1), T'_{s-1}(1), T'_{s-2}(1) \ldots T'_1 (1), T'_0(1) \} \ .
\end{align}
Again, $\nu_n$ is the number of sign changes in $A$, but we now also define $\nu_s$ as the number
of sign changes in $B$.
In this case, the number of zeroes inside the unit circle is $\alpha_n = n - \nu_n$, the number of
zeroes on the unit circle is $\beta_n = 2\nu_s - s$, while the number of zeroes outside of the unit circle is
$\gamma_n = n - \alpha_n - \beta_n$.

For our problem, we analyse the bulk equation, which we write as follows
\begin{equation}
P(x) = \frac{1}{N_2}\bigl(
D_2 (1+x^4) + x^3 (- \lambda t_d - m t_s) + x (\lambda t_d - m t_s) + x^2 (N_2 + 2 m t_s -2 D_2)
\Bigr) \ ,
\end{equation}
with
\begin{align}
N_2 &= \lambda^2 - (m+t_s)^2 &
D_2 &= d_1 d_2 - t_1 t_2 & t_s &= t_1+t_2 & t_d &= t_1-t_2 \ .
\end{align}
The polynomial $P(x)$ is scaled such that $P(1) = 1$.

To proceed, we note that when the system does not exhibit skin effect, the eigenvalues $\lambda$ of the {\em open
chain} that we study lie on the curve given by the eigenvalues of the {\em periodic chain}. We use this information
when analysing the sequences $A$ and $B$ defined above. The eigenvalues for the periodic chain are
given by Eq.~\eqref{eq:lambda-periodic} with $0 \leq k < 2\pi$.

Because the solution in the periodic case satisfies the same bulk equation, we find that $x = e^{i k}$ is a root
of $P(x)$ provided that we set $\lambda = \lambda_+(k)$. Similarly, $x = e^{-i k}$ is a root of $P(x)$ for
$\lambda = \lambda_-(k)$. In both cases, at least one of the roots lies on the unit circle, implying that the
Bistritz algorithm is singular at some level (when $\lambda$ is set to $\lambda_\pm (k)$).
For this reason, we should analyse at which level the algorithm is singular, depending on the parameters of the model.

If we find that the algorithm is singular at level $s$, with $s$ even, we
know that the number of zeroes on the unit circle, given by $\beta_n = 2\nu_s - s$ is also even.
This implies that there are at least two zeroes on the unit circle, because we know that there is at least
one such zero. This in turn implies the absence of the skin effect.
Because the product of the roots $x_1 x_2 x_3 x_4 = 1$, we also know that is it not possible
to have precisely three roots on the unit circle.

\subsection{Skin effect for the model with real parameters.}

In the case of polynomials with complex parameters, the Bistritz algorithm becomes cumbersome, because
constructing the polynomials $T_i (x)$ involves taking the complex conjugate. Therefore, we initially focus on the
case with real parameters $m, t_1, t_2, d_1, d_2$, but of course allow $\lambda$, and hence $N_2$, to be complex.
In this case, we obtain the following results for the polynomials $T_i (x)$.
For $T_4(x)$, we have
\begin{align}
\nonumber
T_4 (x) &=
2 \Bigl( D_2 \Re(N_2) (1-2x^2+x^4) - m t_s \Re(N_2) (x-2x^2+x^3) + i t_d \Im(\lambda N_2^*) (x - x^3) \Bigr) / (N_2 N_2^*) + 2x^2
\\
T_4 (1) &= 2 \\
T_4 (0) &= 2D_2 \Re(N_2) / (N_2 N_2^*) \ . \nonumber
\end{align}
For $T_3(x)$, we have
\begin{align}
\nonumber
T_3 (x) &= 2\Bigl(i D_2 \Im(N_2) (1 + x - x^2 - x^3)  - i  m t_s \Im(N_2) (x-x^2)
- t_d \Re(\lambda N_2^*) (x+x^2) \Bigr)  / (N_2 N_2^*)\\
T_3 (1) &= -4 \Re(\lambda N_2^*) / (N_2 N_2^*)\\
T_3 (0) &= 2 i D_2 \Im (N_2) / (N_2 N_2^*) \ . \nonumber
\end{align}
For $T_2(x)$, we have
\begin{align}
\nonumber
T_2 (x) &= -2 t_d \Re(\lambda) (1-x^2) / (i \Im(N_2)) -2 x\\
T_2 (1) &= -2 \\
T_2 (0) &= -2 t_d \Re(\lambda)/(i \Im(N_2)) \ . \nonumber
\end{align}
Finally, for $T_1(x)$, we have
\begin{align}
\nonumber
N_2 N_2^* \, T_1 (x) &=
2\Bigl(
- D_2 \Im(N_2)^2 / (t_d \Re(\lambda)) + t_d \Re(\lambda N_2^*) \Bigr)(1+x)
+ 2 i m t_s \Im (N_2) (1-x)
\\
N_2 N_2^* \, T_1 (1) &= -4 D_2 \Im(N_2)^2 / (t_d \Re(\lambda)) + 4 t_d\Re(\lambda N_2^*)\\
N_2 N_2^* \, T_1 (0) &= -2D_2 \Im(N_2)^2 / (t_d \Re(\lambda)) + 2 t_d \Re(\lambda N_2^*)
+2 i  m t_s \Im(N_2) \ . \nonumber
\end{align}
We do not explicitly state the constant $T_0$, because it is a long expression and we do not need it for our purposes.

We start by analysing under which conditions $T_3 (x) \equiv 0$. This requires $N_2 = N_2^*$ and
$(\lambda + \lambda^*) t_d =0$.
The first condition $N_2 = N_2^*$ is equivalent to $\Re(\lambda) = 0$ or $\Im(\lambda) = 0$.
The second condition $(\lambda + \lambda^*) t_d =0$ is equivalent to $\Re(\lambda) = 0$ or $t_d = 0$.
Combined, we find that $T_3 (x) \equiv 0$ requires that either $\Re(\lambda) = 0$ or that $\Im(\lambda) = t_d = 0$.

We remark that $\Re(\lambda) = 0$ implies that $4 d_1 d_2 \sin(k)^2 + (m + (t_1+t_2)\cos(k) )^2 \leq 0$,
which can only occur if $d_1$ and $d_2$ have opposite signs (or when $d_1 d_2 = 0$).
On the other hand, if $\Im(\lambda) = 0$ and $t_d = 0$ we need $4 d_1 d_2 \sin(k)^2 + (m + (t_1+t_2)\cos(k) )^2 \geq 0$.

Finally, we note that the explicit form of $\lambda_\pm(k)$ implies that for $t_d = 0$, we have that
either $\Re(\lambda) = 0$ or $\Im (\lambda) = 0$. Hence, for $t_d = 0$ the algorithm is singular at
level $s=4$ (we recall that we assumed that all the parameters of the model are real). This finishes the
analysis of the conditions $T_3 (x) \equiv 0$.

The other way in which we can have two roots on the unit circle is when the algorithm is singular at
level 2, that is when $T_1 (x) \equiv 0$ (and $T_2(0) \neq 0$).
By analysing the form of $T_1 (x)$, making use of the explicit form of the eigenvalues in the
periodic case $\lambda_\pm (k)$, one finds that $T_1 (x) \propto m t_s t_d$.
Because $T_2(0) \neq 0$ for $m=0$ or $t_s = 0$, we obtain that the algoritm is also singular
when either $m=0$ or $t_s = 0$.

\begin{table}[t]
\begin{tabular}{c | c}
condition on parameters & condition on $k$\\
\hline
$m = 0$ & $\forall k$ \\
$t_1 = t_2$ & $\forall k$ \\
$t_2 = -t_2$ & $\forall k$ \\
$d_1 d_2 < 0$ & $4 d_1 d_2 \sin(k)^2 + (m + (t_1+t_2)\cos(k) )^2 < 0 $
\end{tabular}
\caption{Absence of the skin effect, in the case of real parameters}
\label{tab:no-skin-effect-real}
\end{table}

We summarise the result in table \ref{tab:no-skin-effect-real}. For the model with real parameters, there is no skin effect when either
$m = 0$, or when $t_1 = \pm t_2$. In addition, there is no skin effect when $\Re(\lambda) = 0$, which
occurs (over an extended range for $k$) when $4 d_1 d_2 \sin(k)^2 + (m + (t_1+t_2)\cos(k) )^2 < 0$, requiring $d_1 d_2  < 0$ (we note that
$|d_1 d_2|$ should be sufficiently large in order to have $4 d_1 d_2 \sin(k)^2 + (m + (t_1+t_2)\cos(k) )^2 < 0$).
We note that the eigenvalues being real ($\Im(\lambda) = 0$) alone does
not imply that the skin effect is absent. Real eigenvalues can have skin effect when $t_1 \neq t_2$.

To illustrate these results, we plot the (complex) spectrum of the model for several parameters. In all the plots, we use
the following colour conventions. The gray curves represent the eigenvalues of the model with {\em periodic} boundary
conditions. The green dots represent the eigenvalues of the open chain of finite length with $L = 100$ sites.
The blue lines represent eigenvalues of the infinite open chain, corresponding to eigenstates that {\em do} have skin effect.
Finally, the red lines represent eigenvalues of the infinite open chain, corresponding to eigenstates that {\em do not}
have skin effect. The latter eigenvalues also correspond to eigenvalues of the periodic chain.
\begin{figure}[b]
\includegraphics[width=0.8\textwidth]{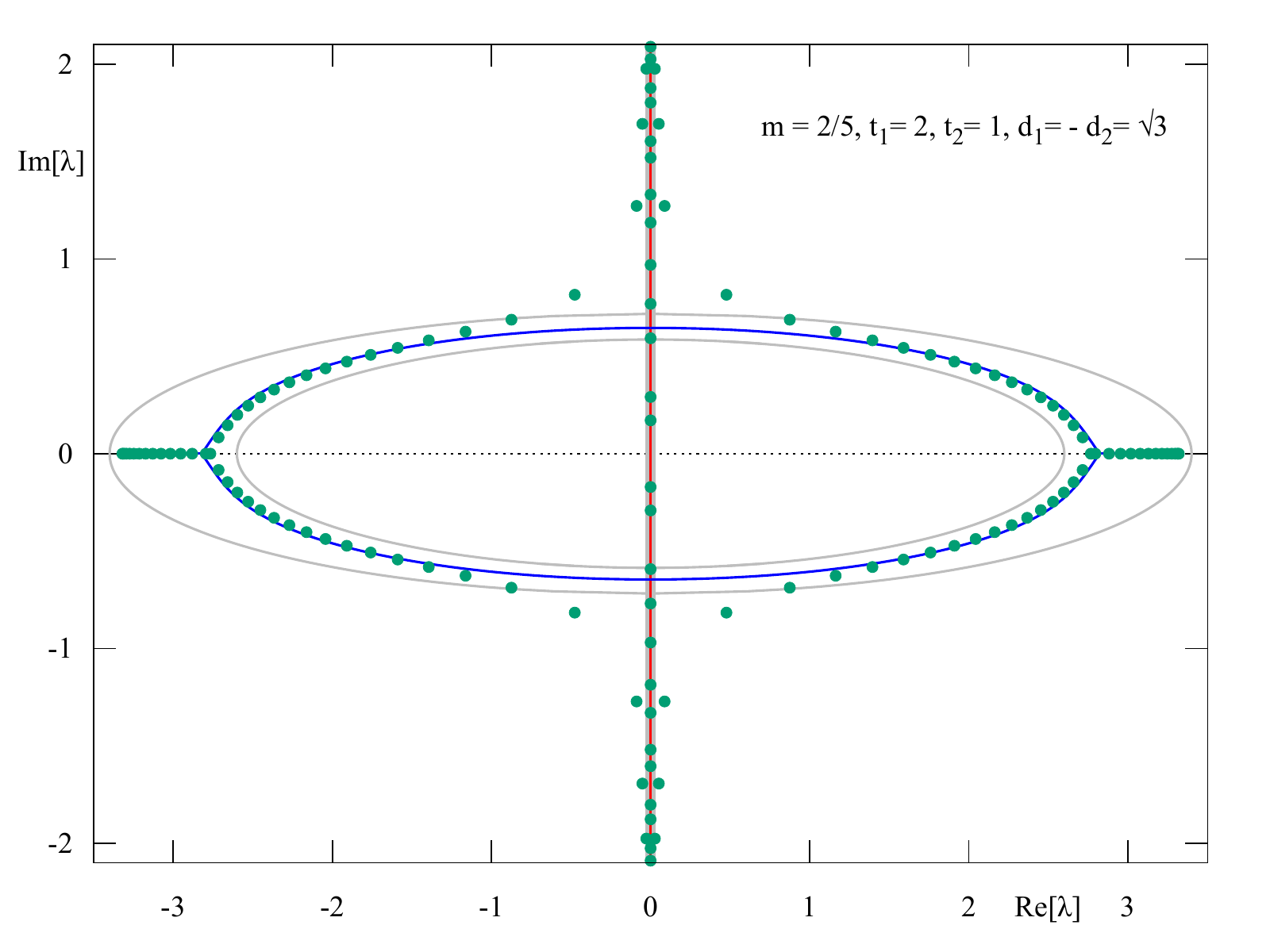}
\caption{The eigenvalues of the model with parameters $m = 2/5$, $t_1 = 2$, $t_2 = 1$, $d_1 = - d_2 = \sqrt{3}$.
The gray lines correspond to the periodic case; the green dots correspond to the open finite chain with $L = 100$;
the blue (red) lines correspond eigenvalues of the open infinite chain whose eigenstates do (do not) exhibit the
skin effect. The eigenvalues along the imaginary axis extent to $\lambda \approx \pm 4.4495 i$.}
\label{fig:case1}
\end{figure}

\begin{figure}[t]
\includegraphics[width=0.49\textwidth]{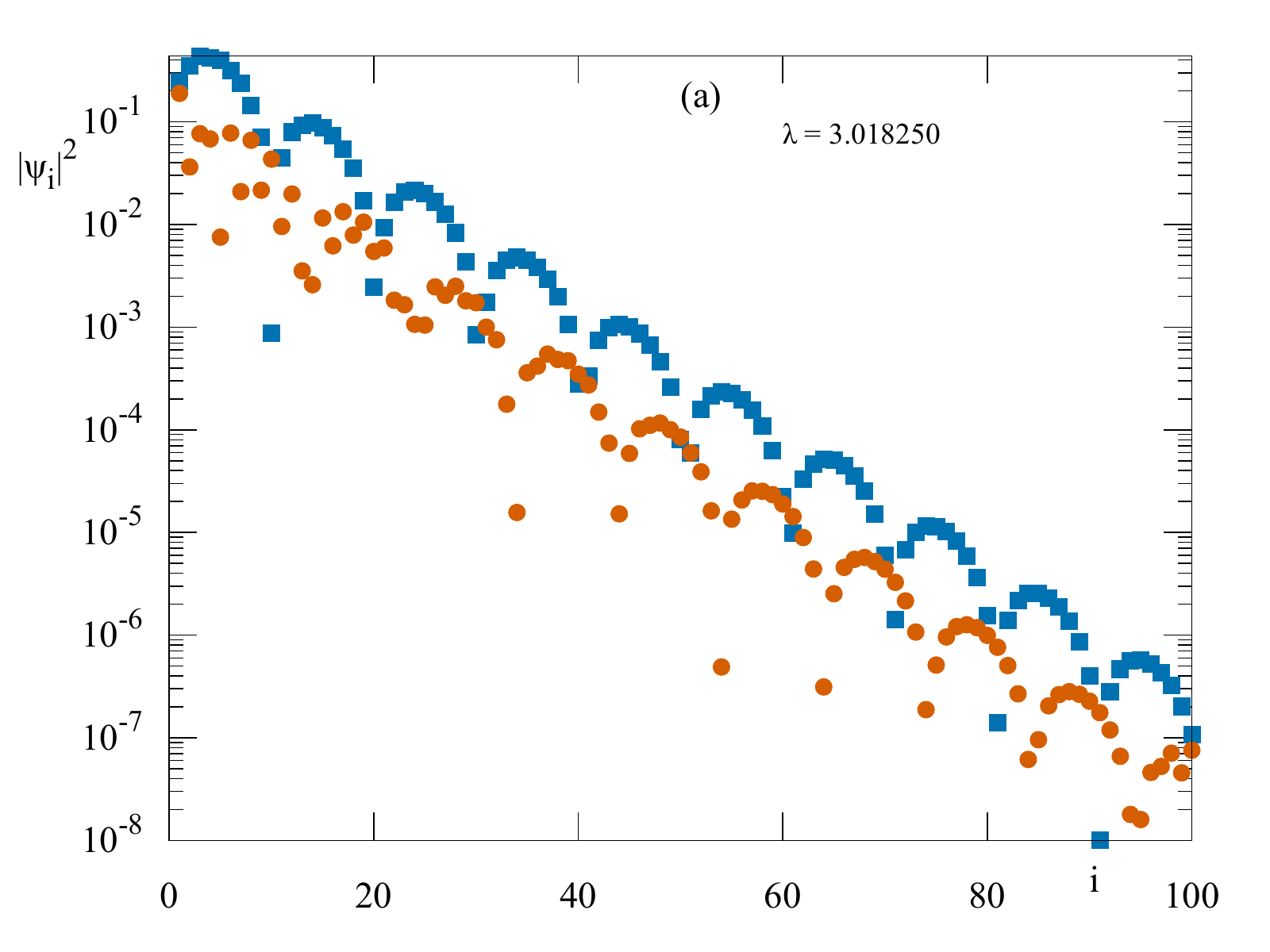}
\includegraphics[width=0.49\textwidth]{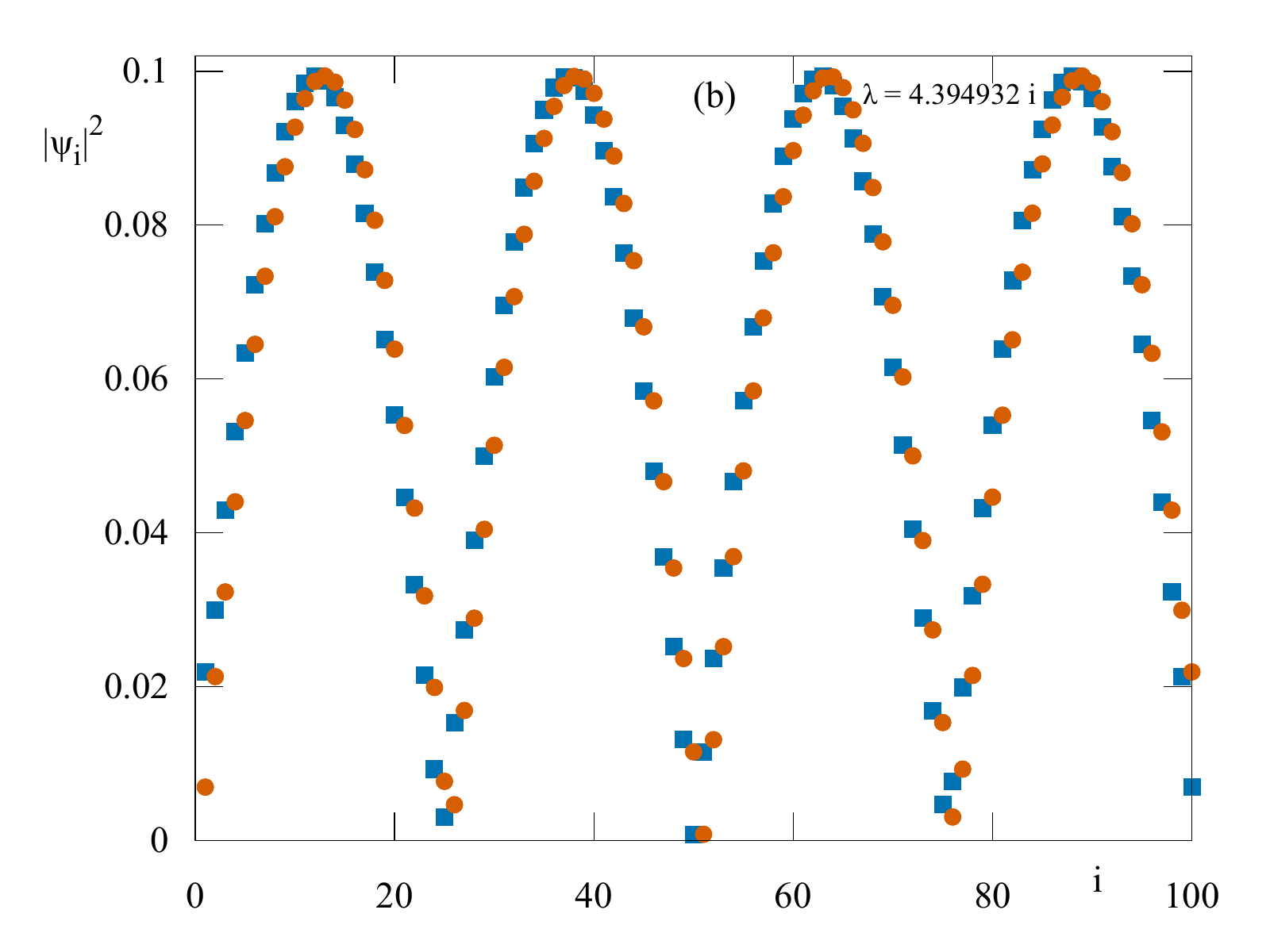}
\caption{The amplitude of eigenstate coefficients $|\psi_i |^2$
as a function of the site index $i$ for two states of the finite $L=100$ model with parameters
$m = 2/5$, $t_1 = 2$, $t_2 = 1$, $d_1 = - d_2 = \sqrt{3}$.
The blue squares (orange dots) correspond to the first (second) component of the eigenvector of a given site $i$.
The left panel (a) corresponds to a real eigenvalue ($\lambda \approx 3.0182$) showing skin effect (note the logarithmic scale),
the right panel (b) corresponds to a purely imaginary eigenvalue ($\lambda \approx 4.3949 i$) without skin effect.
 }
\label{fig:skin-effect}
\end{figure}

\begin{figure}[b]
\includegraphics[width=0.49\textwidth]{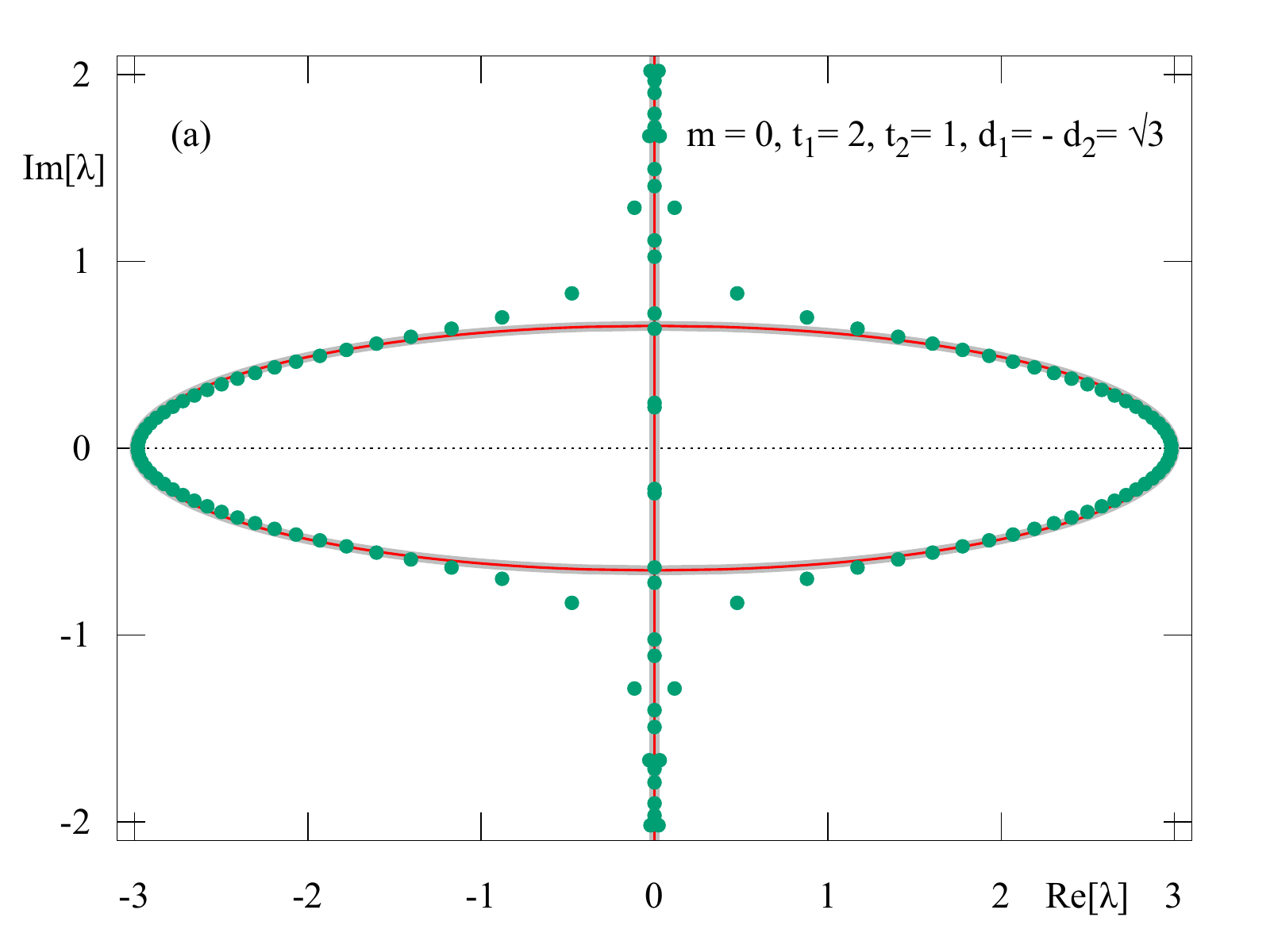}
\includegraphics[width=0.49\textwidth]{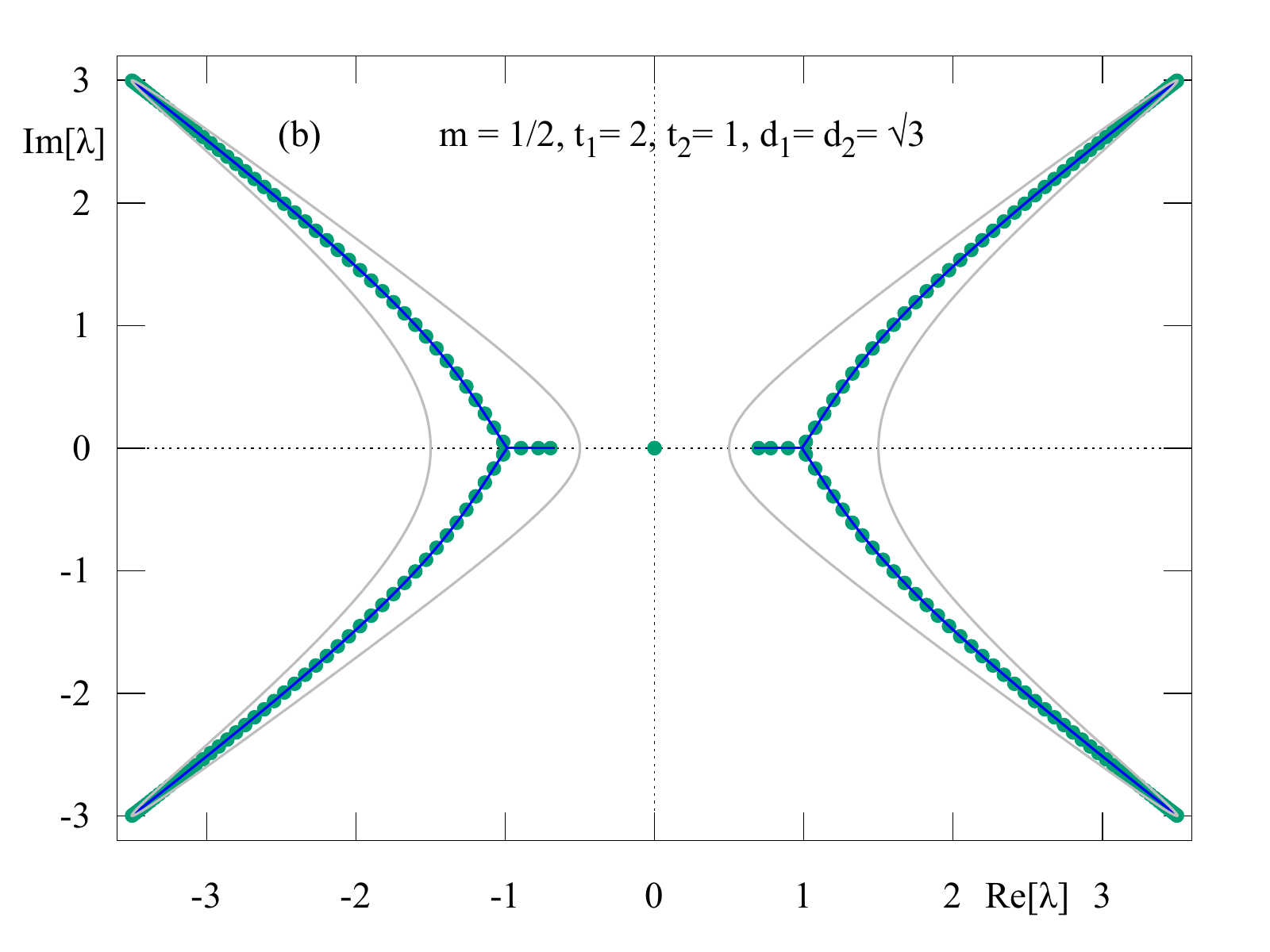}
\caption{The eigenvalues for two sets of parameters. The colour coding is the same as in Fig.~\ref{fig:case1}.
The parameters for the left panel (a) are
$m = 0$, $t_1 = 2$, $t_2 = 1$, $d_1 = - d_2 = \sqrt{3}$;
in this case, none of the eigenstates exhibit skin effect, thus the red and gray curves coincide.
The purely imaginary eigenvalues extend to $\pm (1+2\sqrt{3}) i \approx \pm 4.4641 i$.
We cut off the figure for clarity.
The parameters for the right panel (b) are
$m = 1/2$, $t_1 = 2$, $t_2 = 1$, $d_1 = d_2 = \sqrt{3}$;
in this case, all the eigenstate exhibit skin effect. We note that in this case, there is zero-mode.
 }
\label{fig:case2and3}
\end{figure}

In Fig.~\ref{fig:case1}, we plot the eigenvalues of the system with parameters $m = 2/5$, $t_1 = 2$, $t_2 = 1$, $d_1 = - d_2 = \sqrt{3}$.
The figure clearly shows that for these parameters (which do no fall in one of the classes $m = 0$ or
$t_1 = \pm t_2$), some of the eigenstates of the infinite open chain do show skin effect, while others do not.  
The eigenvalues that are purely imaginary do not show skin effect, while those that lie in the region bounded by
the two ovals (corresponding to eigenvalues of the periodic case) do show skin effect. Interestingly, there are
real (non-zero) eigenvalues that correspond to states that do exhibit skin effect. In the hermitian case, non-zero
(and necessarily real) eigenvalues do not exhibit skin effect. In Fig.~\ref{fig:skin-effect}, we plot the absolute value of the
eigenstate coefficients as a function of position for two eigenvalues of the finite chain namely
$\lambda \approx 3.0182$ in the left panel (a) (real eigenvalue with skin effect, using a logarithmic scale) and
$\lambda \approx 4.3949 i$ in the right panel (b) (purely imaginary eigenvalue without skin effect).
In particular, in the left panel of Fig.~\ref{fig:skin-effect}, we find that the structure of the eigenvector is an
exponentially damped oscillation. The exponential decay signifies the presence of the skin effect.
In terms of the non-Bloch band theory of non-hermitian systems introduced in \cite{YoMu2019}, the eigenvectors
that do not exhibit skin effect have a $\beta$ parameter such that $|\beta| = 1$ (or a real momentum), while the
eigenvectors that do exhibit skin effect have a $\beta$ parameter such that $|\beta| \neq 1$ (i.e., its `momentum'
is complex).

Depending on the parameters, it can also happen that either all the eigenstates of the model have skin effect
(this is the generic case), or none of the eigenstates have skin effect. In Fig.~\ref{fig:case2and3}, we show an
example of either case.

\subsection{Skin effect for the model with complex parameters.}

In this section, we consider the model with complex parameters, generalising the results of the
previous section. As in the case of real parameters, we do not try to find all isolated points for which
the eigenstates do not show a skin effect (this could occur when the curves describing the eigenvalues
in the periodic case, i.e., $\lambda_\pm (k)$, self intersect).

In the previous section, we obtained that $m = 0$, $t_1 = t_2$ and $t_1 = - t_2$ are three
different sufficient conditions implying that there is no skin effect for all the eigenstates in the case of real parameters.
We now argue that these three conditions remain sufficient in the case of complex parameters.

To show this, we consider the Vieta Eqs.~\eqref{eq:vieta}, and solve them for $\kappa$, $s$ and $\lambda$ under the
conditions $m = 0$, $t_1 = -t_2$ or $t_1 = t_2$. When either $m = 0$ or $t_1 = -t_2$, we find that there are solutions
with $\kappa = \pm i$. When $t_1 = t_2$, there are solutions with $\kappa = \pm 1$. Because in all these cases, $|\kappa| = 1$
we find that the ordering of the roots satisfies $|x_1| \geq |x_2| = |x_3| \geq |x_4|$, which shows that the obtained solutions
correspond to actual eigenvalues of the model. Because $| \kappa | = 1$, the eigenstates do not exhibit a skin effect. Indeed,
in these cases the form of $\lambda$ as obtained from the Vieta equations corresponds to the eigenvalues of the periodic case,
$\lambda_\pm (k)$, which has to be true in the absence of the skin effect.

We continue by generalising the results for real parameters, that originated from the condition $T_3 (x) \equiv 0$. In order to do
this, we need the form of the Bistritz polynomials for complex parameters. Because these are quite involved,
we state them in Appendix~\ref{app:bistritz-complex}. In particular, we need $T_3(x)$ as given in Eq.~\eqref{eq:t3-complex}.

We find that the condition $T_3 (x) \equiv 0$ is equivalent to the following conditions on
the parameters in the model
\begin{align}
\Im (D_2 N_2^*) & =  0
&
\Im (m t_s N_2^*) & =  0
&
\Re (\lambda t_d N_2^*) & =  0 \ .
\end{align}
Let us denote the argument of $m$, $t_s$, etc. by $\phi_{m}$, $\phi_{t_s}$, etc., then the conditions above reduce to
\begin{align}
\label{eq:phase-relations}
\phi_{D_2} &= \phi_{N_2} \bmod \pi &
\phi_{m} + \phi_{t_s} &= \phi_{N_2} \bmod \pi &
\phi_{\lambda} + \phi_{t_d} &= \phi_{N_2} + \pi/2 \bmod \pi \ .
\end{align}
We note that $N_2$ is not independent of the parameters. Making use of the explicit form of $\lambda$ in the periodic case, we obtain
\begin{equation}
N_2 = \lambda^2 - (m+t_s)^2 = 2 m t_s \bigl( \cos (k) -1 \bigr) + 4 D_2 \sin (k)^2 + 2 i \lambda t_d \sin (k) \ ,
\end{equation}
which implies that the relations Eq.~\eqref{eq:phase-relations} are satisfied, provided that
$
\phi_{D_2} = \phi_{m} + \phi_{t_s} =  \phi_{\lambda} + \phi_{t_d} + \pi/2 \bmod \pi
$.
To continue, we assume that $\phi_{m} = \phi_{D_2} - \phi_{t_s} \bmod \pi$ and $\phi_{\lambda} = \phi_{D_2} - \phi_{t_d} - \pi/2 \bmod \pi$.
Then, the condition $T_3 (x) \equiv 0$ reduces to
\begin{equation}
| \lambda | ^2 \sin(\phi_{D_2} - 2 \phi_{t_d}) + (|m|^2 - |t_s|^2) \sin(\phi_{D_2} - 2 \phi_{t_s}) = 0 \ .
\end{equation} 
Because we are interested in extended regions in $k$ for which there is no skin effect,  
we obtain that $\phi_{t_d} = \phi_{t_s} \bmod \pi/2$, and
$\phi_{D_2} = 2 \phi_{t_s}  = 2 \phi_{t_d} \bmod \pi$.
Because of the relation $D_2 = d_1 d_2 - t_1 t_2$, we can replace the relation
$\phi_{D_2} = 2 \phi_{t_s}  \bmod \pi$ by
$\phi_{d_1 d_2} = 2 \phi_{t_s} \bmod \pi$, where $\phi_{d_1 d_2}$ is the phase of $d_1 d_2$.

Combined, we find the following conditions, which are necessary in order that the Bistritz algorithm is
singular at level 4
\begin{align}
\phi_m &= \phi_{t_s} \bmod \pi &
\phi_\lambda &= \phi_{t_d} + \pi/2 \bmod \pi &
\phi_{d_1 d_2} &= 2\phi_{t_s} \bmod \pi &
\phi_{t_d} &= \phi_{t_s} \bmod \pi/2 \ .
\end{align}
We still need to check when these conditions are compatible with the explicit form of the eigenvalues in
the periodic case, $\lambda_\pm(k)$ as given in Eq.~\eqref{eq:lambda-periodic}.
By analysing the form of $\lambda_\pm(k)$, taking the phase relations into account, one finds that
the form of $d_1 d_2$ is crucial. Generically, we write $d_1 d_2 = |d_1 d_2 | e^{i \phi_{d_1 d_2}}$.
In Table~\ref{tab:no-skin-effect}, we state the conditions such that the eigenstates do not show a skin effect
(due to $T_3 (x) \equiv 0$), resulting from this analysis.
We note that the third line also follows from the analysis of the case with real parameters in the previous
subsection and the result that phase-rotating the parameters of the model according
to Eq.~\eqref{eq:paramters-phase-rotation} leads to a rigid phase rotation of the spectrum.

\begin{table}[t]
\begin{tabular}{c | c | c | c | c | c}
$t_s$ & $t_d$ & $m$ & $d_1 d_2$ & $\lambda$ & absence skin effect \\
\hline
$ |t_s| e^{i \phi_{t_s}} $ & $\pm i |t_d| e^{i \phi_{t_s}}$ & $\pm |m| e^{i \phi_{t_s}}$ & $+ | d_1 d_2 | e^{2i \phi_{t_s}} $ & $\pm |\lambda | e^{i \phi_{t_s}}$ & $\forall k$ \\ 
$ |t_s| e^{i \phi_{t_s}} $ & $\pm i |t_d| e^{i \phi_{t_s}}$ & $\pm |m| e^{i \phi_{t_s}}$ & $- | d_1 d_2 | e^{2i \phi_{t_s}} $ & $\pm |\lambda| e^{i \phi_{t_s}}$ & $ -4 | d_1 d_2 | \sin(k)^2 + (\pm |m| + |t_s| \cos(k))^2 > 0 $ \\ 
$ |t_s| e^{i \phi_{t_s}} $ & $\pm |t_d| e^{i \phi_{t_s}}$ & $\pm |m| e^{i \phi_{t_s}}$ & $- | d_1 d_2 | e^{2i \phi_{t_s}} $ & $\pm i |\lambda| e^{i \phi_{t_s}}$ & $ -4 | d_1 d_2 | \sin(k)^2 + (\pm |m| + |t_s| \cos(k))^2 < 0 $ \\ 
\end{tabular}
\caption{Sufficient conditions for absence of the skin effect (due to $T_3 (x) \equiv 0$).}
\label{tab:no-skin-effect}
\end{table}

An obvious check on these results is to consider the hermitian case, with $N_2, \lambda, m, D_2, t_s$ all real, and $t_d$ purely imaginary.
In this case, one finds that indeed $T_3 (x) \equiv 0$ implying that the Bistritz algorithm is singular at level 4.
This in turn implies that the eigenstates do not have a skin effect, as expected.

The other way in which the skin effect is absent, is when $T_1 (x) \equiv 0$. The coefficients of the polynomial
$T_1(x)$ are much more involved, see Eq.~\eqref{eq:t1-complex}. We therefore do not attempt to fully characterise for which (complex) parameters of the
model one has $T_1 (x) \equiv 0$. However, above we argued based on the Vieta equations that for either $m=0$, $t_s = 0$, or $t_d = 0$, 
the eigenstates do not show a skin effect. This means that under these conditions $T_1(x) \equiv 0$
even when the other parameters are complex. We are interested in generic results, that is, extended regions of the
curves of eigenvalues, for which the skin effect is absent. We believe that the conditions provided, exhaust all these
cases. The argument in favour of this statement is that we need that $T_1 (x) \equiv 0$
with $\lambda = \lambda_\pm (k)$, for an extended range of $k$. Due to the form of $\lambda_\pm (k)$, this only seems
possible when the various terms of $\lambda_\pm (k)$ have the same argument, or when one or more of the parameters is zero.

\begin{figure}[tb]
\includegraphics[width=0.8\textwidth]{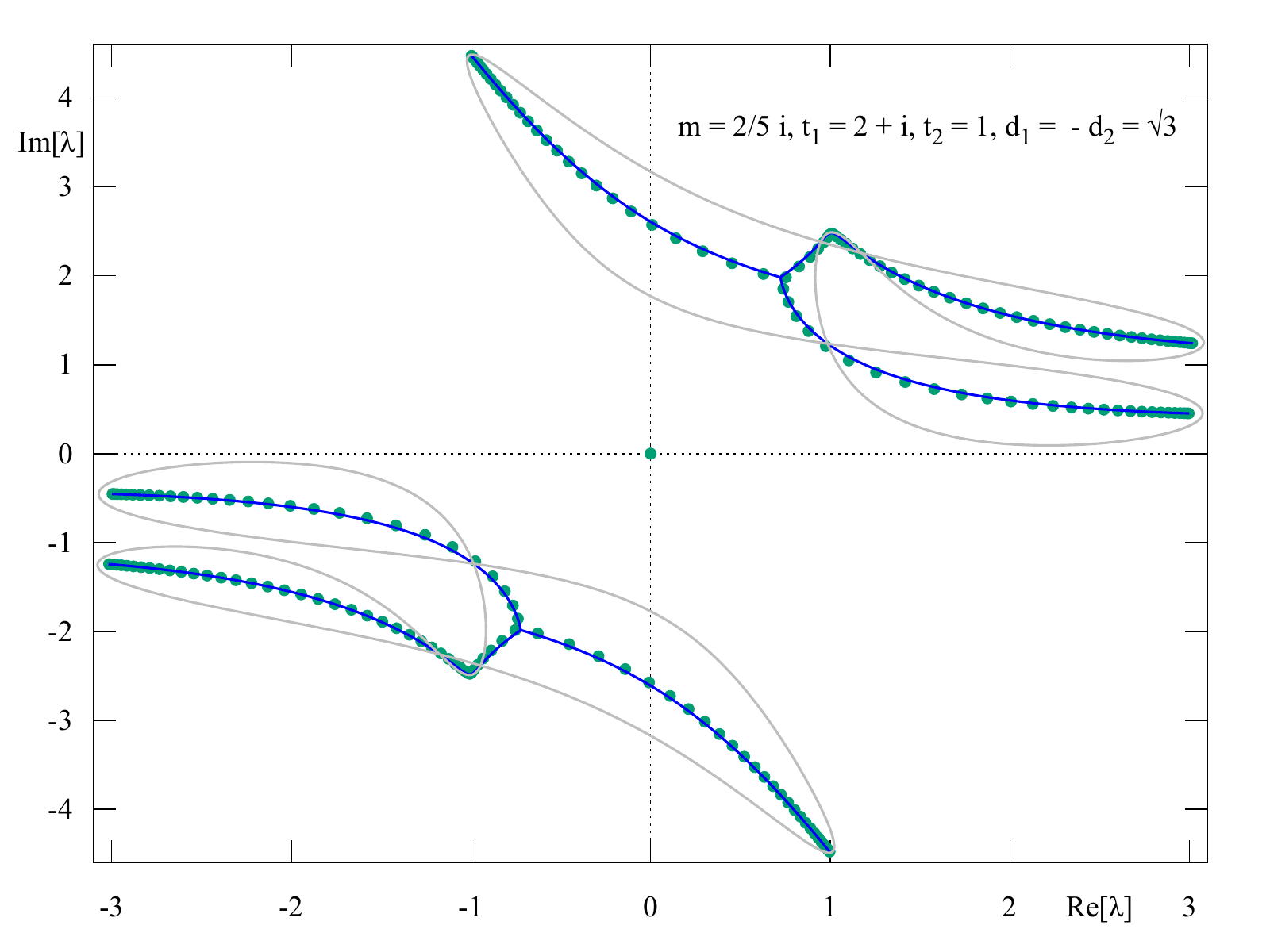}
\caption{The eigenvalues for a set including complex parameters, namely $m = 2i/5$, $t_1 = 2+i$, $t_2 = 1$, $d_1 = -d_2 = \sqrt{3}$.
The colour coding is the same as in Fig.~\ref{fig:case1}.
}
\label{fig:plot-complex}
\end{figure} 
We conclude this section with a characteristic example of the eigenvalues for a case with complex
parameters, namely $m = 2i/5$, $t_1 = 2+i$, $t_2 = 1$, $d_1 = -d_2 = \sqrt{3}$ as shown in Fig.~\ref{fig:plot-complex}.
In this generic case, all the generic eigenstates exhibit the skin effect. As expected, the spectrum is significantly more complex compared
to the cases we showed with real parameters. We note that the gray curves of the eigenvalues $\lambda_\pm(k)$ in the periodic case
intersect themselves. The eigenvalues of the open chain in the large system size limit (given by the blue curves) cross these intersection
points. This means that the eigenstates corresponding to these (six) special eigenvalues do not have a skin effect.
We checked this behaviour explicitly, by solving the bulk Eq.~\eqref{eq:bulk-x}, confirming that two solutions for $x$ indeed have
modulus one. In addition, we checked that the Bistritz polynomial $T_1 (x) \equiv 0$ for the given parameters and the eigenvalue $\lambda$.

\section{Analysis of an example}
\label{sec:summary}

\begin{figure}[t]
\includegraphics[width=0.8\textwidth]{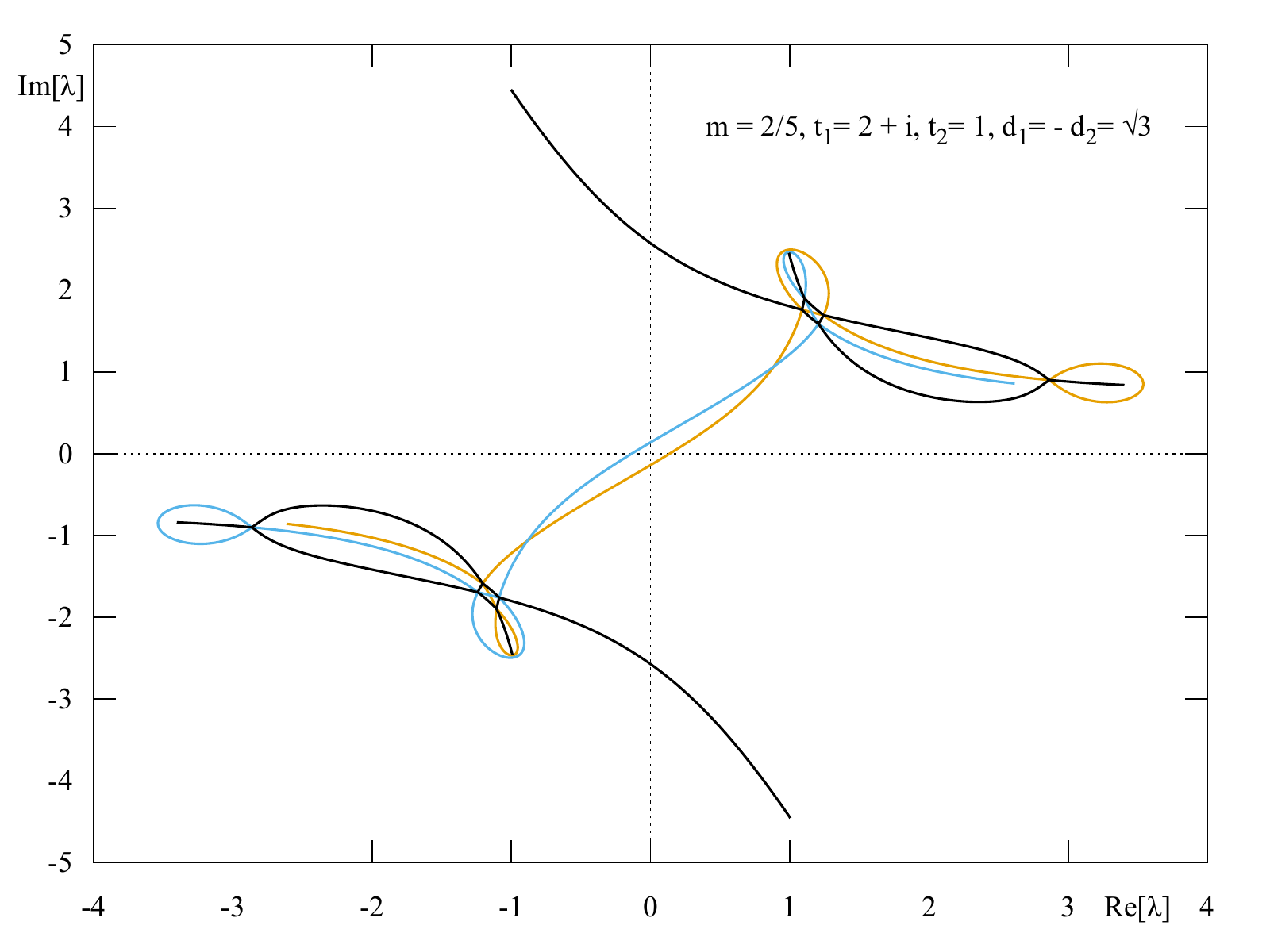}
\caption{Solutions for $\lambda$ of the Vieta equations, for $m = 2/5$, $t_1 = 2+i$, $t_2 = 1$, $d_1 = -d_2 = \sqrt{3}$.
Only the black lines correspond to actual eigenvalues. For the blue and dark-yellow lines, the solutions $x_i$ of the bulk
equation correspond to one of the branches that do not correspond to actual eigenvalues.}
\label{fig:discussion-1}
\end{figure}

We studied the non-hermitian Kitaev chain for general complex parameters, pushing analytical methods as far
as possible. We now summarise our results, by means of an example. We use the chain with parameters
$m = 2/5$, $t_1 = 2+i$, $t_2 = 1$, $d_1 = -d_2 = \sqrt{3}$ for this purpose.
One of the main results we obtained in this paper, is the characterisation of the eigenvalues of the
infinite size system, in terms of three Vieta Eqs. \eqref{eq:vieta}, for $\lambda$ which we repeat here for convenience,
\begin{align}
&\bigl( \kappa + \frac{1}{\kappa}\bigr) \bigl[ (s + \frac{1}{s}) + 2 \cos\alpha \bigr] = \frac{2 m (t_1 + t_2)}{d_1 d_2 - t_1 t_2} \nonumber\\
&\bigl( \kappa - \frac{1}{\kappa}\bigr) \bigl[ (s + \frac{1}{s}) - 2 \cos\alpha \bigr] = \frac{2 \lambda (t_1 - t_2)}{d_1 d_2 - t_1 t_2} \nonumber\\
&( \kappa + \frac{1}{\kappa})^2 + 2 \cos\alpha (s + \frac{1}{s}) = \frac{\lambda^2 - m^2 - (t_1 + t_2)^2}{d_1 d_2 - t_1 t_2} \ .
\end{align}
where $\kappa$ and $s$ are in general complex.
The eigenvalue curves $\lambda$ are obtained by varying $0 \leq \alpha < 2 \pi$, and obtaining the
solution of the three Vieta Eqs. \eqref{eq:vieta}. A priori, the solutions of these equations are continuous
and smooth as a function of the parameter $\alpha$. Importantly, only those $\lambda$ that correspond
to solutions that satisfy $1/|s| \leq |\kappa|^2 \leq |s|$ are actual eigenvalues
of the model as explained in Sec.~\ref{sec:Linfinity}. This explains the observed branched structure of the
actual eigenvalues.

In Fig.~\ref{fig:discussion-1}, we show the eigenvalues of the model with parameters
$m = 2/5$, $t_1 = 2+i$, $t_2 = 1$, $d_1 = -d_2 = \sqrt{3}$ as an illustration. The actual
eigenvalues (the black lines) form a rather intricate pattern, which can be explained in terms of the three
different branches of solutions of the Vieta equations. Fig.~\ref{fig:discussion-1} also shows
the other two branches (the blue and yellow lines), that do not correspond to eigenvalues of the
model. It is interesting to note the regions where the black lines `intersect'. Here, the eigenvalues
do not cross, nor do they repel, but form a rhombic structure.

\begin{figure}[ht]
\includegraphics[width=1.0\textwidth]{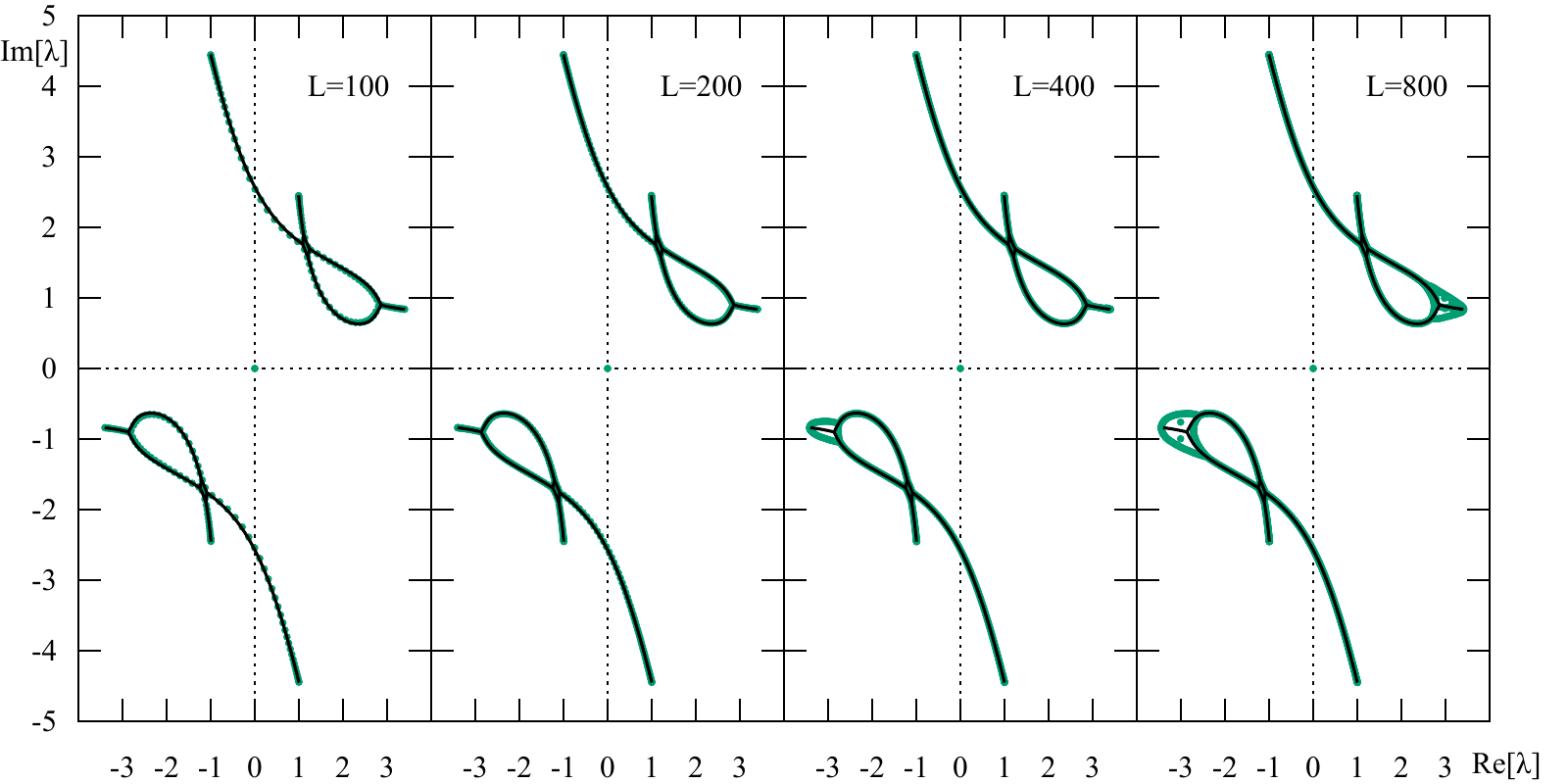}
\caption{Finite size eigenvalues $\lambda$ for the model with parameters
$m = 2/5$, $t_1 = 2+i$, $t_2 = 1$, $d_1 = -d_2 = \sqrt{3}$.
The black lines correspond to actual eigenvalues in the limit $L\rightarrow\infty$.
The different panels show the numerically obtained eigenvalues (using machine precision) for
the sizes $L=100$, $L=200$, $L=400$ and $L=800$, showing the instability of the algorithm
(green dots).
}
\label{fig:discussion-2}
\end{figure}

We saw that having a concise characterisation of the eigenvalues of the infinite system size explains
the structure of the curves the eigenvalues lie on. 
There is another reason why having a concise characterisation for the infinite system size model is useful,
apart from that its interesting in its own right.
Namely, as is well known, obtaining the eigenvalues for large, non-hermitian systems showing skin-effect is often
numerically unstable. We illustrate this using the same example, by plotting the eigenvalues,
as obtained using {\em machine precision} diagonalisation for system sizes $L = 100$, 
$L=200$, $L=400$ and $L=800$. The results are shown in Fig.~\ref{fig:discussion-2} as the
green dots, together with the infinite system size results. 

We clearly see that for sizes $L=100$ and $L=200$, the eigenvalues closely follow the
black lines corresponding to the infinite size model, with only minor difference due to the
finite size effects. However, already for $L=400$, there
is a region where the finite size eigenvalues deviate substantially from curves for infinite
size. Moreover, the `eigenvalues' in this region do not satisfy particle-hole symmetry (dictating
that if $\lambda$ is an eigenvalue, so is $-\lambda$), which clearly indicates that these values
obtained by the diagonalisation algorithm are incorrect. For $L=800$, the situation gets worse.
In principle, one can obtain the correct eigenvalues even for these larger system sizes, if one
uses a diagonalisation algorithm employing higher precision arithmetic, but in practise, this will
be much slower that obtaining the infinite system size results. 

By making use of the exact solution, we studied the presence of the skin effect.
There exist methods to determine if a model exhibits skin effect for a given set of
parameters. Here, we also characterise which eigenvalues have eigenstates
showing the skin effect, or rather, which ones do not show skin effect.
We formulated this condition in terms of the solutions
for $x$ of the bulk Eq.~\eqref{eq:bulk-gen}, namely if at least two (out of the four)
solutions lie on the unit circle, the corresponding eigenstate does not have a
skin effect. We used the Bistritz algorithm to determine under which conditions
there are states without skin effect, and to which eigenvalues these correspond.
We provided a full characterisation for the non-hermitian model with real parameters.
For the general model with complex parameters we provide sufficient conditions
for the absence of the skin effect, which we believe are also necessary.

Finally, we studied under which conditions, the non-hermitian Kitaev chain has a zero mode
(for the hermitian Kitaev chain, this corresponds to the region where the model is in the
topological phase). It turns out that this region has perhaps a more complicated structure than one
would expect. Namely, the model has a zero mode when the following condition is satisfied,
\begin{equation}
|\Im (\arccos \bigl( -\frac{m}{2 \sqrt{t_1 t_2 - d_1 d_2}} \bigr)) |  <  |\Im (\arccos\bigl( \frac{t_1 + t_2 }{2 \sqrt{t_1 t_2 - d_1 d_2}}  \bigr))| \ .
\end{equation}
At the boundary of this region, the model is gapless, but even outside of this region,
the model can be gapless. In the hermitian version of the model, this would correspond to a metallic,
gapless system.

\section{Discussion}
\label{sec:discussion}
 
We studied the non-hermitian Kitaev model for arbitrary, complex parameters. By using a novel method,
we obtained a concise characterisation of the eigenvalues in the thermodynamic limit for arbitrary complex
parameters. Using this method,
we explained the branched structure of the eigenvalues in the complex plane. In addition, we used the
solution to obtain for which parameters, the model exhibit a skin effect, and if so, to which eigenvalues this
pertains. For real parameters, we obtained this in full, while for arbitrary complex parameters, we obtained
sufficient conditions, which we believe are also necessary. We fully characterised the parameters for which
the model exhibits a zero mode. Finally, we discussed the stability issues that arise when one tries to
numerically obtain the eigenvalues for large systems.

 Despite the fact that non-hermitian one-dimensional models have been studied in great detail,
it would be interesting to apply our method to characterise the eigenvalues in the thermodynamic
limit to other, more complicated systems. One direct extension of the model we consider here, is
to add terms that interpolate between open and periodic boundary conditions. Considering different
models, one can think of systens with larger unit cells and/or longer range interactions.
Obviously, this will lead to higher order equations and more complicated expressions.

\appendix

\section{Bistritz polynomials for complex parameters}
\label{app:bistritz-complex}

In this appendix, we give the Bistritz polynomials $T_i (x)$ (with $4\geq i \geq 1$) in the general case, that is
for complex parameters. These results generalise the Bistritz polynomials given in Sec.~\ref{sec:skin-effect} for real
parameters (and complex eigenvalues $\lambda$). We refer to that section for more details. More details
on the Bistritz algorithm can be found in \cite{Bi2002}.

For arbitrary complex parameters and eigenvalue $\lambda$, the polynomial $T_4 (x)$ reads
\begin{align}
T_4(x) &=  2\Bigl( \Re(D_2 N_2^*) (1 - 2 x^2 + x^4) + i \Im(\lambda t_d N_2^*) (x - x^3)
- \Re(m t_s N_2^*) (x - 2 x^2 + x^3) \Bigr) / (N_2 N_2^*) + 2  x^2
\nonumber \\
T_4(1) &= 2 \\
T_4(0) &= 2 \Re(D_2 N_2^*)  / (N_2 N_2^*) \ . \nonumber
\end{align}
For $T_3(x)$, we obtain the following expression
\begin{align}
\label{eq:t3-complex}
T_3 (x) &= 2\Bigl( -i \Im(D_2 N_2^*) (1 + x - x^2 - x^3) + i \Im(m t_s N_2^*) (x-x^2)
- \Re(\lambda t_d N_2^*) (x+x^2) \Bigr)  / (N_2 N_2^*)
\nonumber \\
T_3 (1) &= -4 \Re(\lambda t_d N_2^*) / (N_2 N_2^*)\\
T_3 (0) &= -2 i \Im(D_2 N_2^*) / (N_2 N_2^*) \ . \nonumber
\end{align}
$T_2(x)$ is given by 
\begin{align}
T_2 (x) &=
2\Bigl(
\Re(\lambda t_d D_2^*) (1 - x^2) - i \Im(m t_s D_2^*) (1 - 2 x + x^2)\Bigr) / (i \Im(D_2 N_2^*)) - 2 x
\nonumber \\
T_2 (1) &= -2 \\
T_2 (0) &= 2\Bigl(\Re(\lambda t_d D_2^*) - i \Im(m t_s D_2^*)\Bigr) / (i \Im (D_2 N_2^*)) \ . \nonumber
\end{align}

Finally, $T_1 (x)$ is given by
\begin{align}
\label{eq:t1-complex}
&N_2 N_2^* T_1 (1) = 2 (1+x) \Bigl[
\Re(\lambda t_d N_2^*) - \frac{\Im(N_2 D_2^*)^2\Re(\lambda t_d D_2^*)}{\Re(\lambda t_d D_2^*)^2 + \Im (m t_s D_2^*)^2}
\Bigr] \\
\nonumber &
- 2 i (1-x) \Bigl[
\Im(m t_s N_2^*) + 4 \Im(N_2 D_2^*) + \frac{\Im(N_2 D_2^*)^2\Im(m t_s D_2^*)}{\Re(\lambda t_d D_2^*)^2 + \Im (m t_s D_2^*)^2}
- 4 \frac{\Im(N_2 D_2^*)\Re(\lambda t_d D_2^*)^2}{\Re(\lambda t_d D_2^*)^2 + \Im (m t_s D_2^*)^2}
\Bigr] \ ,
\end{align}
resulting in the following expression for $T_1 (1)$, 
\begin{equation}
T_1 (1) = \frac{4}{N_2 N_2^*} \Bigl(
\Re (\lambda t_d N_2^*) - \frac{\Im (D_2^* N_2)^2 \Re(\lambda t_d D_2^*)}{\Re(\lambda t_d D_2^*)^2 + \Im (m t_s D_2^*)^2} 
\Bigr) \ . 
\end{equation}
Though we do not need it, we give an expression for the constant $T_0$, in terms of $T_2(x)$ an $T_1(x)$, for completeness
\begin{equation}
T_0 = 2 \Re( T_2(0) / T_1(0)) T_1 (1) - T_2 (1) \ .
\end{equation}

\end{document}